\documentclass{aa}  

\usepackage{graphicx}
\usepackage{txfonts}
\begin{document}

   \title{An underlying universal pattern in galaxy halo magnetic fields}

   \subtitle{}

   \author{Ioannis Myserlis
          \inst{1,2}
          \and
          Ioannis Contopoulos\inst{3,4}}%\fnmsep\thanks{}}
   %\authorrunning{Myserlis \& Contopoulos} % for "referee" mode

   \institute{Instituto de Radioastronom\'ia Milim\'etrica, Avenida Divina Pastora 7, Local 20, E-18012 Granada, Spain\\
              \email{imyserlis@iram.es}
         \and
             Max-Planck-Intitut f\"ur Radioastronomie, Auf dem H\"ugel 69, D-53121 Bonn, Germany
         \and
             Research Center for Astronomy and Appl. Math., Academy of Athens, Athens 11527, Greece\\
             \email{icontop@academyofathens.gr}
         \and
             National Research Nuclear University (MEPhI), Moscow 115409, Russia
             %\thanks{}
             }

   \date{\today}%Received September 15, 1996; accepted March 16, 1997}

% \abstract{}{}{}{}{} 
% 5 {} token are mandatory
 
  \abstract
  % context heading (optional)
  % {} leave it empty if necessary  
  % aims heading (mandatory)
  % {}
  % methods heading (mandatory)
  % {}
  % results heading (mandatory)
  % {}
  % conclusions heading (optional), leave it empty if necessary 
  % {}
  {Magnetic fields in galaxy halos are in general very difficult to observe. Most recently, the CHANG-ES collaboration (Continuum HAlos in Nearby Galaxies – an EVLA Survey) investigated in detail the radio halos of 35 nearby edge-on spiral galaxies and detected large scale magnetic fields in 16 of them. We used the CHANG-ES radio polarization data to create Rotation Measure (RM) maps for all galaxies in the sample and stack them with the aim to amplify any underlying universal toroidal magnetic field pattern in the halo above and below the disk of the galaxy. We discovered a large-scale magnetic field in the central region of the stacked galaxy profile, attributable to an axial electric current that universally outflows from the center both above and below the plane of the disk. A similar symmetry-breaking has also been observed in astrophysical jets but never before in galaxy halos. This is an indication that galaxy halo magnetic fields are probably not generated by pure ideal magnetohydrodynamic (MHD) processes in the central regions of galaxies. One such promising physical mechanism is the Cosmic Battery operating in the innermost accretion disk around the central supermassive black hole. We anticipate that our discovery will stimulate a more general discussion on the origin of astrophysical magnetic fields.}


   \maketitle
%
%-------------------------------------------------------------------

\section{Introduction}
\label{sec:intro}

We are entering the era of new high-resolution, broadband radio facilities
%(e.g., LOw-Frequency ARray -- LOFAR, new generation Very Large Array -- ngVLA, Atacama Large Millimeter Array -- ALMA, Square Kilometre Array -- SKA, etc.)
(e.g., LOFAR, ngVLA, ALMA, SKA, etc.)
that will allow us to investigate in unprecedented detail the distribution and properties of cosmic magnetic fields along the line of sight, toward and within astronomical objects. Large-scale, coherent galactic magnetic fields play a fundamental role in star formation, cosmic ray propagation, and the dynamics of astrophysical accretion disks as well as galactic and extragalactic jets \citep{Kronberg2016}. Their origin and evolution remain one of the key open questions of modern astrophysics.

The distribution of cosmic magnetic fields along the line of sight can be investigated using linearly polarized radiation. The complex intensity of linear polarization $\Pi$ is defined as:
\begin{equation}
\Pi\equiv Q+iU=Pe^{i2\chi}=mIe^{i2\chi} 
\end{equation}                          
where, $I, Q$, and $U$ are the observed Stokes parameters, $P$ is the modulus of $\Pi, \chi$ is the observed Electric Vector Position Angle (EVPA), and $m\equiv \sqrt{Q^2+U^2}/I$ is the observed degree of linear polarization. Faraday rotation changes the observed orientation of the EVPA when polarized light is propagated through a magnetoionic medium from an initial value $\chi_0$ at position $r_0$ to an observed value $\chi$ via the following relation:
\begin{equation}
\label{eq:Faraday Rotation}
\chi=\chi_0+{\rm RM}\cdot\lambda^2 
\end{equation}    
where $\lambda$ is the wavelength and the Faraday Rotation Measure (hereafter RM) is defined as: 
\begin{equation}
{\rm RM}\equiv \frac{e^3}{2\pi m_e^2 c^4 } \int_{r_0}^0 n_e B_{||} {\rm d}r 
\end{equation}
($r$, $n_e$, $B_{||}$ are the distance, electron number density and line-of-sight magnetic field of the magnetoionic medium respectively, and $r_0$ is the distance of the source of polarized radiation). The amplitude and sense of rotation probe the strength and orientation of the magnetic field component along the line of sight. The dependence of the polarization angle on wavelength $\lambda$ becomes more complex than Eq.~(\ref{eq:Faraday Rotation}) if the source of the polarized radiation is the magnetoionic medium itself (i.e., internal Faraday rotation).

In this paper, we further investigate a very interesting discovery of the past decade, namely the existence of a universal pattern in the large-scale magnetic fields associated with astrophysical black holes. In a series of papers \citep{Gabuzda2012,Christodoulou2016}, Contopoulos and collaborators presented evidence for a universal toroidal magnetic field structure in astrophysical jets, namely that (when present) electric currents along kpc-scale astrophysical jets always flow outwards. Similar large-scale magnetic field patterns have been observed across the outer layers of galactic disks using background polarized radio sources. Two characteristic examples of this phenomena are found (i) in the extended region around the radio jet of Centaurus A \citep{Feain2009}, and (ii) in our own Galaxy \citep{Pshirkov2011,Oppermann2015}. In both cases, the inferred large scale axial electric currents also flow outwards.

There are several reasons why it has up to now been very hard to observe such a universal pattern in galactic halos. Firstly, the magnetoionic medium in the halo is very tenuous and can only be observed in the radio by stacking together images of tens of galaxies \citep[e.g. figure 6 in][]{Wiegert2015}. Secondly, detailed studies of line-of-sight magnetic fields are better performed by means of Faraday RM synthesis \citep{Brentjens2005} which requires broadband radio observations with good spectral resolution (i.e. a large and well-sampled wavelength range). Such data have only been sparsely available so far. Furthermore, galactic winds, supernova explosions, crossings of the galactic disk by neighboring galaxies, astrophysical jets, etc., severely disrupt any underlying universal magnetic field structure in the halo. 

In this paper, we set out to investigate the direction of the large scale axial electric current flowing away from or toward the center of a typical disk galaxy seen edge-on, by studying the large scale toroidal magnetic field above and below the galactic disk. We decided to use data from the CHANG-ES collaboration (Continuum HAlos in Nearby Galaxies - an EVLA Survey). The CHANG-ES sample consists of 35 edge-on spiral galaxies observed with the Karl G. Jansky Very Large Array (VLA) in wide bands centered at 1.5 GHz and 6 GHz \citep{Irwin2012}. The original goal of that survey was to provide new insight into the halo and disk-halo activity in spiral galaxies. The physical conditions in these regions were probed in considerably more detail than ever before, and the observations suggest that most galaxies in the sample experience outflowing winds. Several galaxies in the sample contain AGNs revealed mostly through circular polarization \citep{Irwin2018}. Bipolar X-shaped magnetic field structures can be seen when stacking the linear polarization maps of all galaxies \citep{Krause2020}. Such structures are otherwise masked in most of the individual galaxy maps. The polarization data together with a rotation measure synthesis analysis have revealed large scale coherent magnetic field structures that have never been seen before in any galaxy, such as regularly reversing magnetic fields on kpc scales both in the disk and the halo, likely generated by magnetic dynamo action in the disk \citep{Krause2020,Mora-Partiarroyo2019}. The CHANG-ES project is described in \citet{Irwin2012} and details about its first data release, relevant to our present analysis are provided in \citet{Wiegert2015}.

%We work with the publicly available data \citep{Wiegert2015} and look for systematic features in the RM map around the center of the stacked edge-on disk galaxy, indicative of a coherent axial electric current in the halo.

This paper is organized as follows: in Section~\ref{sec:RM analysis} we analyze the publicly available CHANG-ES data \citep{Wiegert2015} to create RM maps for all galaxies in the sample. In Section~\ref{sec:RM_profile} we stack the RM maps to reveal an underlying universal toroidal magnetic field structure, indicative of a coherent axial electric current in the halo. Finally, in Section~\ref{sec:discussion} we discuss our results in the framework of the Cosmic Battery mechanism for the origin of astrophysical magnetic fields in the disk surrounding an astrophysical black hole \citep{Contopoulos1998}.

%--------------------------------------------------------------------
\section{RM analysis of CHANG-ES data}
\label{sec:RM analysis}

In this section we describe the procedure that we follow to obtain RM values from the publicly available CHANG-ES data. As input, we use the total intensity (Stokes $I$), linear polarization and polarization angle $\chi$ (EVPA) maps from the D configuration of VLA. In particular, we make use of the EVPA maps at both the L and C bands, centered at 1.5~GHz and 6~GHz respectively, to calculate the RM between them. The corresponding FITS files were obtained from the CHANG-ES data release website \footnote{\url{https://www.queensu.ca/changes}}.
For our analysis, we use the images made with uniform uv-weighting (robust=0) and corrected for the primary beam, labeled as ``Rob 0'' and ``PBcor'' in the data release web site, respectively. Nevertheless, we repeated our analysis with the gaussian uv-tapered version of the maps (labeled as ``UVtap'') and/or the non-primary beam corrected maps (``no PBcor'') and found that the following results do not change significantly.

To construct the RM maps, we compare the EVPA images in the L and C bands (hereafter L-map and C-map) for each galaxy in the sample on a pixel-by-pixel basis. Since the resolution between the two images is different, we start with the L-map which has a pixel size 2-5 times larger than that of the C-map. For each pixel in the L-map that the EVPA value $\chi_L$ was not flagged \footnote{The images are flagged wherever Stokes $Q$ and $U$ fall below 3~sigma.}, we select all non-flagged pixels in the C-map that fit into its area and calculate their average value $\overline{\chi}_C$. We note here that averaging EVPA measurements may lead to errors due to their $180^\circ$ ambiguity \footnote{For example, the mean of $90^\circ$ and $-90^\circ$, which corresponds to the same polarization angle, is $0^\circ$, that is perpendicular to the direction of the averaged EVPA measurements.}. To avoid this, we calculate the Stokes $Q$ and $U$ parameters that correspond to the selected C-map pixels using the linear polarization and polarization angle $\chi$ (EVPA) maps and then we calculate the average EVPA as:
\begin{equation}
\overline{\chi}_C=\frac{1}{2}  \arctan\left(\frac{\overline{U}}{\overline{Q}}\right)\equiv\frac{1}{2}  \arctan\left(\frac{\sum_{i=1}^N U_i}{\sum_{i=1}^N Q_i }\right)
\end{equation}
where $N$ is the number of selected C-map pixels\footnote{During this process we exclude C-map pixels with linear polarization degree less than 0.5\%, as suggested by the data release website.}. We found that this method delivers more stable results between neighboring pixels than averaging the EVPA measurements directly.

The result of the previous process is two maps for each source, both at the resolution of its L-map with each applicable pixel having a pair of EVPA measurements ($\chi_L, \overline{\chi}_C$) at 1.5~GHz and 6~GHz, respectively, that can be used to obtain an RM value as:
\begin{equation}
\label{eq:RM}
{\rm RM}=\frac{\chi_L-\overline{\chi} _C}{\lambda_L^2-\lambda_C^2} \end{equation}
where $\lambda_L$ and $\lambda_C$ are the L and C band wavelengths respectively.

As with any other polarization angle measurement, both $\chi_L$ and $\overline{\chi}_C$ have an ambiguity of $180^\circ$. This means that, in principle, RM in Eq.~(\ref{eq:RM}) can have an infinite number of values obtained by adding or subtracting integer values of $180^\circ$ from either $\chi_L$ or $\overline{\chi}_C$.
For consistency, we decided to minimize the difference between $\chi_L$ and $\overline{\chi}_C$ and consider the value of RM in Eq.~(\ref{eq:RM}) that is minimum in absolute terms. For the minimization, we use $\overline{\chi}_C$ as the pivot EVPA measurement, since the higher-frequency C-band data are less affected by Faraday rotation.

We note that the limited frequency coverage of the input data as well as the RM minimization method described above restricts the RM calculation to a maximum absolute value of about $50\ {\rm rad/m}^2$. Therefore, our RM results may be underestimated, especially in regions with high electron density  or magnetic field strength. Nevertheless, we decided to follow the RM minimization approach to avoid extreme differences between neighboring pixels in the RM maps of individual galaxies. Our results suggest that the RM minimization can mitigate (extreme) differences in the range of 75-105 ${\rm rad/m}^2$ between neighboring pixels by about 60\%. As a representative example, in the map of galaxy NGC\,4666 two neighboring pixels have RM values of $+39.9\ {\rm rad/m}^2$ and $-52.1\ {\rm rad/m}^2$ initially and after the application of the RM minimization these values become $+39.9\ {\rm rad/m}^2$ and $+43.1\ {\rm rad/m}^2$. Since these are adjacent pixels, a small rather than a large RM difference is expected. A more robust estimate of the RM may be obtained with EVPA maps at additional wavelengths.
Nevertheless, we note that the minimization was implemented only in about 16\% of all pixels and we found that our results do not change significantly even if these pixels are completely excluded from the analysis.
%\textbf{Nevertheless, it is interesting that despite the restricted RM range imposed by the RM minimization method, we detect a universal pattern in galaxy halo magnetic fields that persists after random $180^\circ$ rotations of individual galaxies (Sect.~\ref{sec:RM_profile}).}

Finally, the RM maps were corrected for the contribution from the Galactic Faraday depth using the estimate of figure 15 in \citet{Oppermann2015}. The Galactic RM contribution was calculated as the average of several pixels around the location of each galaxy in the ``Galactic foreground'' map obtained from the publication website \footnote{\url{https://wwwmpa.mpa-garching.mpg.de/ift/faraday/2014/index.html}}. The resulting Galactic RM values were subtracted from the corresponding RM maps of all galaxies that we analyzed, except NGC\,2613 where the Galactic RM was found to be about $174\ {\rm rad/m}^2$, which is outside the range that can be probed using the RM calculation methodology described above. We would also like to note that polarization angles have been corrected for ionospheric effects by the CHANG-ES team \citep{Wiegert2015}. The final RM maps for most galaxies in the sample are presented in Appendix~\ref{app:RMmaps}.

\section{Evidence for a universal pattern via RM stacking}
\label{sec:RM_profile}

We investigate a magnetic field pattern in the halo above and below the galactic disk that does not depend on the orientation of the galactic disk (up or down) and is thus universal. The RM maps created individually for each galaxy through the procedure described in Sect.~\ref{sec:RM analysis} were found to cover only small parts of the galaxy (see Appendix~\ref{app:RMmaps}). This makes it difficult to probe the large-scale magnetic field. To address this issue, we decided to stack together all of our RM maps.

The first step required for RM stacking is to align the galaxies.
To do that we obtain their coordinates and position angles from \citet{Irwin2012} and the HyperLeda catalog \citep{Makarov2014}. 
First, we set the $l$ (galactic longitude) and $b$ (galactic latitude) axes origin of the RM maps to the center of each galaxy, using the coordinates listed in \citet{Irwin2012}. Then, we rotate each map so that the galactic plane is along the horizontal direction, using
% their position angles listed in the HyperLeda catalog \citep{Makarov2014}
%% removed to reduce text:
the transformation:
\begin{eqnarray}
\label{eq:alignment}
l & = & \delta {\rm RA} \cdot \sin {\rm PA}+\delta {\rm DEC} \cdot \cos {\rm PA}\nonumber\\ 
b & = & \delta {\rm RA} \cdot \cos {\rm PA}+\delta {\rm DEC} \cdot \sin {\rm PA}
\end{eqnarray}
where $\delta{\rm RA}$ and $\delta{\rm DEC}$ is the distance of each pixel from the center of the galaxy along the right ascension and declination directions, $l$ and $b$ the distance in the reference frame of the given galaxy (longitude and latitude) and $0 \leq {\rm PA} \leq 180^\circ$ is the angle of the major axis of the galactic disk measured clockwise (north to east) from north \citep{Makarov2014}. Obviously, there is a $180^\circ$ ambiguity in this alignment but this is not important in our efforts to measure axial electric currents in galaxies seen edge-on.
Following this step, we inspected the maps and applied minor adjustments ($\leq |6^\circ|$) to achieve better horizontal alignment.

The second step required for RM stacking is to rescale the galaxies so that all $22\ \mu {\rm m}$ infrared galactic disks have the same angular size in the sky \citep[obtained from Col.~4 of Table~6 in][]{Wiegert2015}. With this procedure, very small and very large galaxies are scaled up and down, respectively, so that they all contribute in the same physical scale. The outcome of these two steps is a distribution of RM values in each map according to the distance from the center of each galaxy in units of their $22\ \mu {\rm m}$ infrared disk radius $r_{\rm disk}$, namely ${\rm RM}(x,y)$. We note that five galaxies were excluded from the RM stacking, based on the justification described in Sect.~6 of \citet{Wiegert2015}, namely NGC\,660 and NGC\,4438 (distorted structure), NGC\,4594 and NGC\,5084 (large beam), and UGC\,10288 (map dominated by background AGN). A similar procedure of angular and physical scaling was followed by \citet{Wiegert2015} and \citet{Krause2020} to stack the total flux density and linear polarization maps of the CHANG-ES sample, respectively, and reveal the extent of the (median) galaxy halo and its X-shaped magnetic field structure on the plane of the sky.

\begin{figure*}
 \centering
 \vspace{-0.32cm}
 \includegraphics[trim={40 40 30 10}, clip, width=1.0\textwidth,angle=0.0]{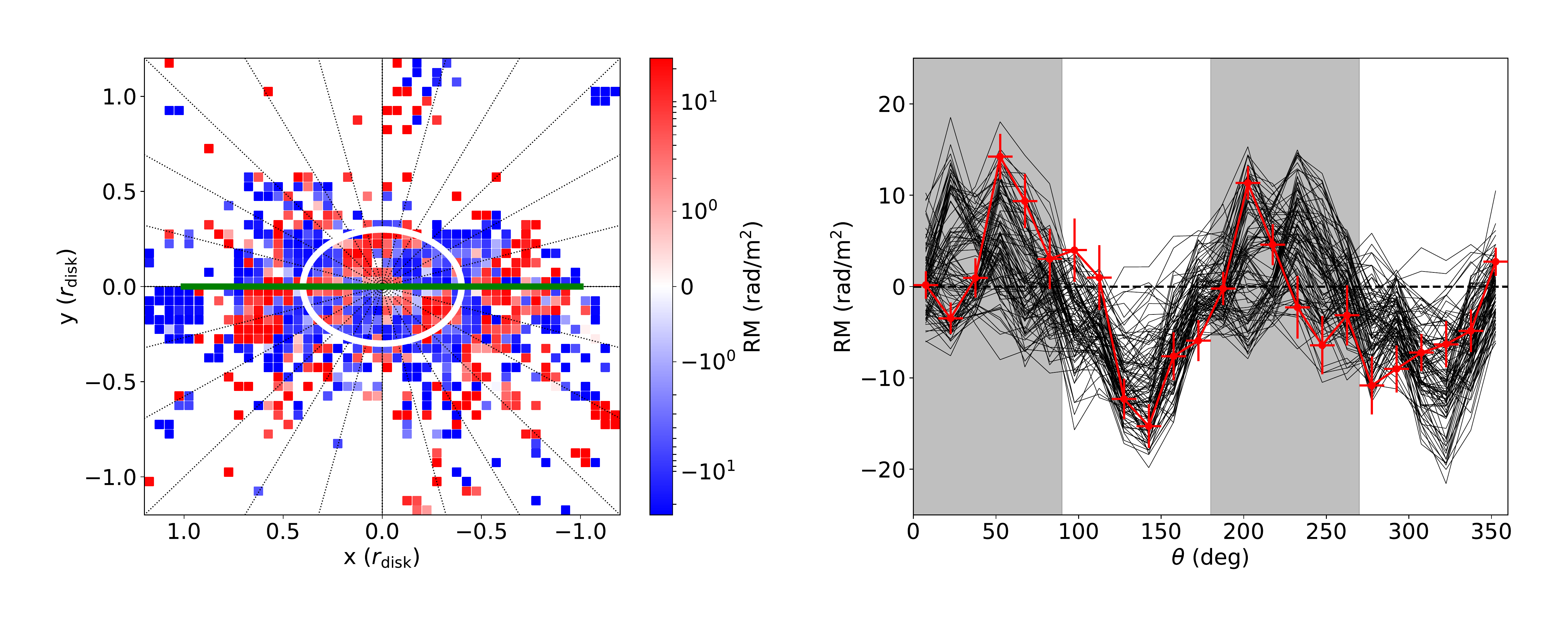}
 \caption{\textbf{Left:} mean RM map corresponding to the red line in the right panel. \emph{Color plot:} binned RM values (logarithmic color scale from -25 to $+25\ {\rm rad/m}^2$). \emph{Thick white line:} elliptical region where most bins contain contributions from several galaxies. \emph{Thick green line:} horizontal extent of edge-on rescaled $22\ \mu {\rm m}$ infrared disk. \emph{Thin dotted lines:} $15^\circ$ angular sectors around the center. \textbf{Right:} \emph{Red line:} mean RM distribution around the center of the mean CHANG-ES galaxy inside the white ellipse on the left (see text for details).
 %, after aligning each individual galaxy according to Eqs.~(\ref{eq:alignment}). 
 Angles measured clockwise from the left horizontal semi-axis. The alternating grey and white colors of the background indicate the four quadrants of the mean RM map. \emph{100 thin black lines:} similar to red line after randomly rotating half of the aligned galaxies by $180^\circ$. The underlying universal pattern survives on average after these random $180^\circ$ rotations.}
 \label{fig:main}
\end{figure*}

To proceed with our analysis, we bin together all RM values of all galaxies into one mean map (Fig.~\ref{fig:main}, left). We then divide the area of the whole map into sectors of $15^\circ$ around the center. In order to %investigate the presence of an underlying magnetic field pattern similar to the one observed in our own Galaxy \citep{Pshirkov2011,Oppermann2015} and 
avoid special patterns dominated by individual galaxies, we decided to focus our analysis in a region of the mean map that contains contributions from several galaxies in each bin. It turns out that this region can be outlined by an ellipse around the center with semi-major and semi-minor axes of $0.4r_{\rm disk}$ and $0.3r_{\rm disk}$, respectively ($r_{\rm disk}$ is the semi-major axis of the rescaled galaxies\footnote{For a direct comparison between the $22\ \mu {\rm m}$ infrared disk size and the extent of the (median) halo in radio, we point the reader to Fig.~6 of \citet{Wiegert2015}}). We thus collect all RM values in each angular sector inside this ellipse (highlighted in white in the left panel of Fig.~\ref{fig:main}). 
The names of all galaxies (24 total) that contribute to the bins inside the white ellipse, along with the number of data points and corresponding percentage, are listed in Table~\ref{tab:data_in_ellipse} and their RM maps are shown in Appendix~\ref{app:RMmaps}. Notice that not all galaxies listed in Table~\ref{tab:data_in_ellipse} contribute to each bin inside the white ellipse. The bins inside the ellipse contain data from 7 galaxies on average, with a maximum of 13 galaxies per bin\footnote{There are a few bins in the ellipse with data from one or two galaxies, but the vast majority of them (94\%, 139/148) contain data from at least 3 galaxies.}, while the ones outside the ellipse contain contributions from only 2 galaxies on average.

Since we are investigating large scale axial electric currents, we are looking for gradients in RM across the vertical axis in both hemispheres (upper and lower) that would manifest the reversal of the line-of-sight magnetic field direction imposed by the toroidal magnetic field configuration. In the right panel of Fig.~\ref{fig:main} we plot the mean RM value, together with the error of the mean \footnote{The error of the mean is defined as $[ \sum_{i=1}^n ({\rm RM}_i-\overline{\rm RM})^2/(n(n-1))]^{1/2}$, where $n$ is the number of ${\rm RM}_i$ measurements and $\overline{\rm RM}$ is the average RM value in each sector.}, in each sector located at an angle $\theta$ starting from the left horizontal semi-axis and increasing clockwise.

%What we obtain is very exciting.
Our results provide strong evidence for \emph{a systematic universal variation of the mean RM near the center of the mean disk galaxy across all four quadrants around the center}. As shown in Fig.~\ref{fig:main} (left), the upper left and lower right quadrants inside the white ellipse have primarily positive RM values (red), which means that the line-of-sight magnetic field points towards the observer, while the upper right and lower left quadrants inside the white ellipse have primarily negative RM values (blue), which means that the line-of-sight magnetic field points away from the observer. This universality disappears outside the drawn ellipse, possibly due to the dominance of randomly oriented large-scale magnetic field components, like the ones discovered by \citet{Krause2020}, from the 1-2 individual galaxies that contribute to these bins. The magnetic field pattern around the center is attributable to a large-scale electric current along the axis of the mean galaxy that flows outwards in both hemispheres (upper and lower). This is consistent with the evidence presented by Contopoulos and collaborators \citep{Gabuzda2012,Christodoulou2016} and the magnetic field pattern observed in our own Galaxy \citep{Pshirkov2011,Oppermann2015}. Our work is complementary to the short study of Lynden-Bell \citep{Lynden-Bell2013} who discovered that the axial magnetic field at the centers of disk galaxies seen face-on lies along the direction of $\Omega$ of the galactic rotation (he called this ``magnetism along spin'').

\begin{table}
\caption{Galaxies that contribute data within the white ellipse of Fig.~\ref{fig:main}. The second and third columns list the number of RM measurements within the ellipse and the corresponding fractions, respectively. The RM maps are shown in Appendix~\ref{app:RMmaps}.} % title of Table
\label{tab:data_in_ellipse}      % is used to refer this table in the text
\centering                          % used for centering table
\begin{tabular}{l r r}        % centered columns (4 columns)
\hline\hline                 % inserts double horizontal lines
Name & data points & percentage  \\    % table heading 
\hline                        % inserts single horizontal line
NGC\,4631   &  343  & 17.8 \%  \\
NGC\,3079   &  200  & 10.4 \%  \\
NGC\,3628   &  199  & 10.3 \%  \\
NGC\,3556   &  177  &  9.2 \%  \\
NGC\,4666   &  173  &  9.0 \%  \\
NGC\,5775   &  150  &  7.8 \%  \\
NGC\,4565   &  107  &  5.5 \%  \\
NGC\,4192   &   96  &  5.0 \%  \\
NGC\,891    &   95  &  4.9 \%  \\
NGC\,4217   &   78  &  4.0 \%  \\
NGC\,3735   &   74  &  3.8 \%  \\
NGC\,4157   &   71  &  3.7 \%  \\
NGC\,4845   &   45  &  2.3 \%  \\
NGC\,3044   &   42  &  2.2 \%  \\
NGC\,5907   &   21  &  1.1 \%  \\
NGC\,2683   &   20  &  1.0 \%  \\
NGC\,4388   &   13  &  0.7 \%  \\
NGC\,3448   &    7  &  0.4 \%  \\
NGC\,4302   &    7  &  0.4 \%  \\
NGC\,4013   &    4  &  0.2 \%  \\
NGC\,3432   &    3  &  0.2 \%  \\
NGC\,4096   &    3  &  0.2 \%  \\
NGC\,5792   &    2  &  0.1 \%  \\
NGC\,5297   &    1  &  0.1 \%  \\
\hline
Total       &  1931 & 100 \%   \\
%NGC\,2613   &    0  &  0.0 \%  \\
%NGC\,2820   &    0  &  0.0 \%  \\
%NGC\,2992   &    0  &  0.0 \%  \\
%NGC\,3003   &    0  &  0.0 \%  \\
%NGC\,3877   &    0  &  0.0 \%  \\
%NGC\,4244   &    0  &  0.0 \%  \\
\hline                                   %inserts single line
\end{tabular}
\end{table}

\begin{figure}
 \centering
 \vspace{-0.32cm}
 \includegraphics[width=0.5\textwidth,angle=0.0]{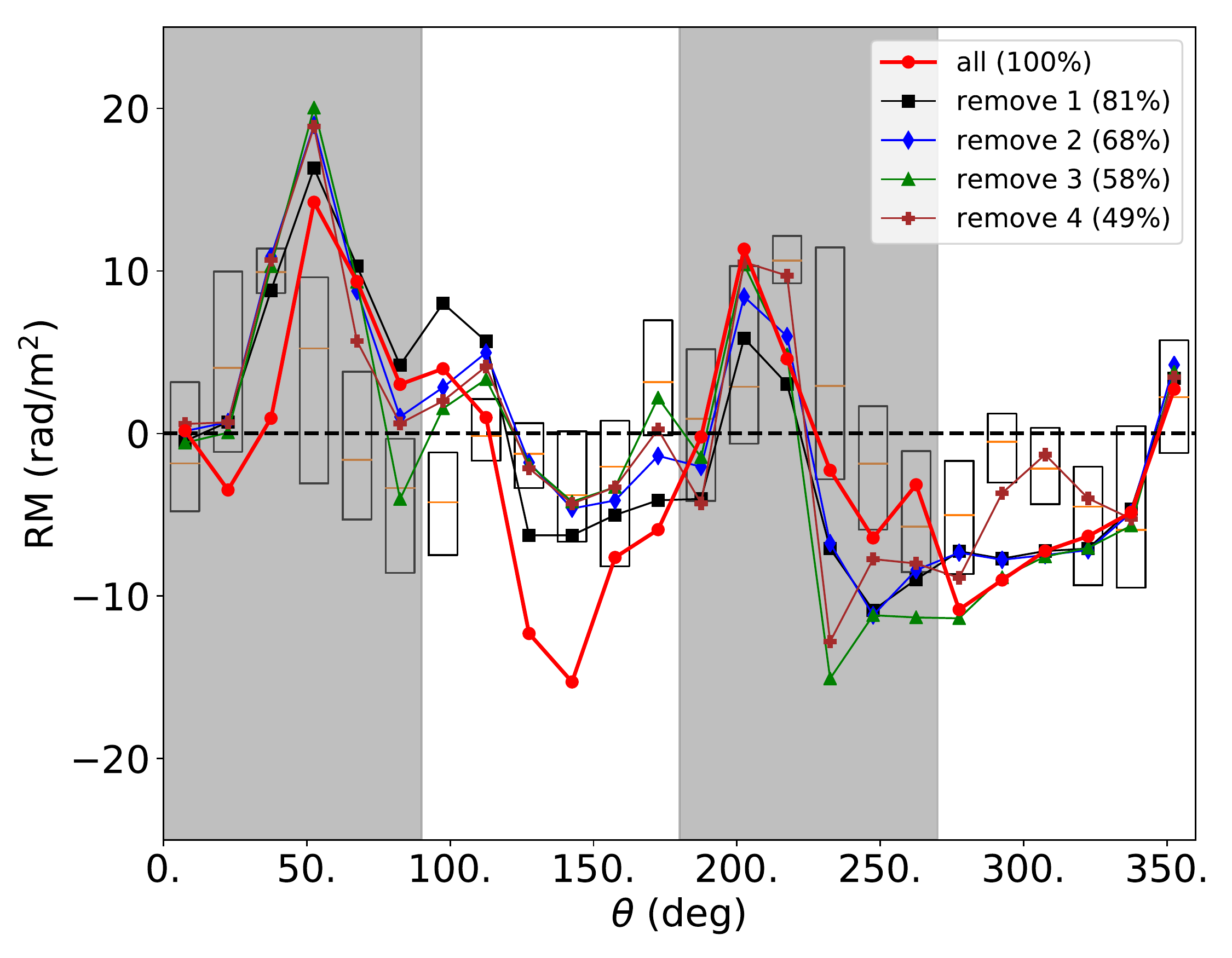}
 \caption{Superposed mean RM angular distributions after removing one-by-one the four galaxies with the most RM values inside the white ellipse of Fig.~\ref{fig:main} (the remaining percentage of RM values shown in the legend). Details of box plot in the text. The universal pattern roughly persists.}
 \label{fig:remove numerous}
\end{figure}

We performed several tests to verify that this is not a random occurrence but is rather a strong indication of an underlying universal pattern in galactic halo magnetic fields. To confirm that the pattern is not due to our particular alignment of individual galaxies (Eqs.~\ref{eq:alignment}), we randomly rotate half of the galaxies upside-down several times and plot their corresponding angular distributions of RM in the right panel of Fig.~\ref{fig:main}. We clearly see that the universal pattern persists. The average spearman $\rho$ correlation coefficient when each line in Fig.~\ref{fig:main} is fitted with a $\sin(2\theta)$ dependence is equal to 0.70 with average p-value $5\times 10^{-3}$, indicating a strong and significant correlation with the quadrupolar RM profile around the center at a $3\sigma$ significance level. We also checked for a possible bias in the pattern from the 4 galaxies with the most RM values inside the white ellipse (about 50\%), namely NGC\,4631, NGC\,3628, NGC\,3079 and NGC\,3556. When we remove them one-by-one from the stacking, the angular RM distribution roughly persists, as seen in
(Fig.~\ref{fig:remove numerous}).
In order to show the stacking uncertainty in that figure, we consider the most extreme case  (remove 4 galaxies that contribute cumulatively 50\% of the RM data inside the white ellipse) and perform 100 random $180^\circ$ rotations to half of the remaining galaxies (similar to the black lines on the left panel of Fig.~\ref{fig:main}). The regions that the resulting lines are passing through are marked with box plots extending to the first and third quartiles of the distribution of points in each angle bin (the median is marked with the horizontal orange line in each box plot). We see that the quadrupolar angular RM distribution roughly persists also in the box plot representation.

\begin{figure}
 \centering
 \vspace{-0.32cm}
 \includegraphics[width=0.5\textwidth,angle=0.0]{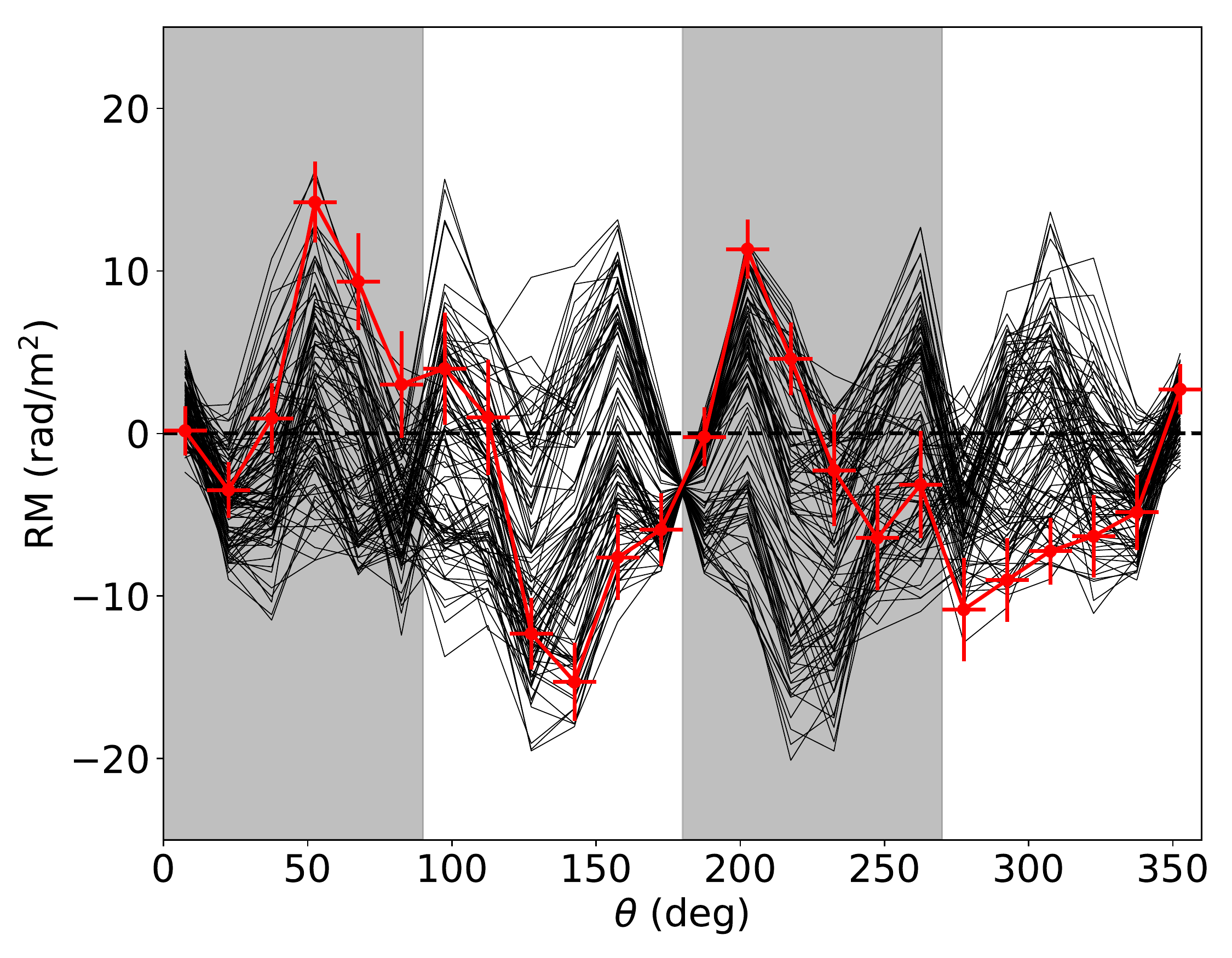}
 \caption{\emph{Thin black lines:} Superposed 100 mean RM angular distributions after randomly mirror-reversing vertically half of the galaxies in the sample. \emph{Red line} as in Fig.~\ref{fig:main}, that is the mean RM angular distribution without mirror-reversing vertically any of the galaxies. The red line is over-plotted to highlight that that no universal pattern is present in the thin black lines.}
 \label{fig:revererseY}
\end{figure}

Finally, the universal pattern that we advocate is anti-symmetric with respect to the galaxy plane, therefore, when we randomly reverse half of the maps with respect to the horizontal axis (i.e. reverse the sign of $y$) and repeat our stacking procedure several times, the RM distribution no longer exhibits an ordered, large scale configuration (Fig.~\ref{fig:revererseY}). In that case, the average spearman $\rho$ correlation coefficient for a $\sin(2\theta)$ fit is equal to -0.001 with average p-value 0.29. 

\section{Discussion and conclusions}
\label{sec:discussion}

In this paper we showed preliminary observational evidence that in all galaxies, the toroidal magnetic field $B_\phi$ configuration in the innermost halo is the same as that observed for our own Galaxy \citep{Pshirkov2011,Oppermann2015}. This common pattern consists a universal symmetry-breaking in the distribution of large scale cosmic magnetic fields which cannot be accounted for by pure ideal magnetohydrodynamic (MHD) processes \footnote{The equations of ideal MHD are quadratic in the magnetic field $B$, thus, a reversal of $B$ everywhere does not affect the dynamics.}, and therefore, a dynamo field advected into the halo cannot account for the effect that we advocate. If true, the detection of a universal symmetry-breaking suggests a relation between the polarity of the large-scale axial magnetic field $B_z$ and the sense of rotation $\Omega$ of the disk. The argument is simple: ``winding'' $B_z$ by the rotation $\Omega$ generates a $B_\phi$ of a certain polarity, and if there is a relation between the polarity of $B_z$ and $\Omega$, reversing the direction of $\Omega$ would reverse both the polarity of $B_z$ and the direction of its winding, which means that it would result in {\it always the same} polarity of $B_\phi$, hence the claim of a \emph{universal symmetry-breaking} \citep[the interested reader may consult Fig.~2 of][]{Christodoulou2016}.

This is supported by the Cosmic Battery mechanism for the origin of astrophysical magnetic fields in the disk surrounding an astrophysical black hole \citep{Contopoulos1998}. According to that mechanism, the aberrated radiation pressure on the electrons of the electron-proton plasma in orbit around the black hole decelerates the electrons and thus induces an azimuthal electric field opposite to the direction of rotation. The rotation (curl) of this electric field generates poloidal magnetic field flux which, near the black hole, is along the direction of disk-hole rotation $\Omega$ (“magnetism along spin”), and in the surrounding disk is opposite to $\Omega$. A toroidal magnetic field component is obtained when the poloidal magnetic field is twisted by differential rotation in the accretion disk around the galaxy's central supermassive black hole. This twisting results in an axial electric current that always flows away from the center \citep[e.g. Fig.~1 in][]{Christodoulou2016}. We propose that the universal magnetic field that we measure is the above toroidal field transported by galactic winds to large distances in the galactic halo. An alternative explanation could be that the Hall current plays a key role in the magnetohydrodynamics of the galactic disk \citep{Konigl2010}. More observations are needed to support our preliminary evidence. 

We present here an order of magnitude estimate: the Cosmic Battery around a central supermassive black hole of a few times $10^8$ solar masses generates a magnetic field $B_{\rm ISCO}\sim 10^3$~G within the radius of the innermost stable circular orbit (ISCO) $r_{\rm ISCO}\sim 10^{14}$~cm \citep{Contopoulos2015}. This is the base of a disk wind which accelerates the flow out to about $r_0\sim 10\ r_{\rm ISCO}\sim 10^{15}$~cm, where the toroidal and poloidal magnetic field components are expected to be of the order $B_0\sim 10^{-2}B_{\rm ISCO}\sim 10$~G. Beyond that distance, the wind coasts without further acceleration, and the poloidal field drops with distance $r$ as $r^{-2}$, whereas the toroidal field drops as $r^{-1}$. Thus, at distances of a few kpc in the galactic halo, the toroidal field in the wind reaches values $B\sim B_0 (r/r_0)^{-1}\sim$ a few $\mu {\rm G}$. Furthermore, for halo thermal electron number density $n_e$ on the order of a few times $10^{-2}\ {\rm cm}^{-3}$ \citep{Cordes2002}, we obtain a rough estimate of the halo RM over a few kpc on the order of $10\ {\rm rad/m}^2$, which are the values obtained in Fig.~\ref{fig:main}.

The stacking of several galactic RM maps that we performed in the present work to enhance a potential underlying universal pattern certainly deserves further investigation as a new method of detecting universal magnetic field patterns in the halo. We acknowledge that this is preliminary work with a small sample of galaxies, and that one should definitely repeat it with a more extended sample. In conclusion, we hope that our discovery will stimulate a systematic search for halo magnetic fields with the upcoming new generation of high-resolution, broadband radio telescopes, and a more general discussion on the origin of astrophysical magnetic fields.

\begin{acknowledgements}
We used the HyperLeda database (\url{http://leda.univ-lyon1.fr}). The authors would like to thank the two anonymous referees, for the careful reading of the manuscript and their constructive comments. The authors also thank N. R. MacDonald, the internal MPIfR referee, for his insightful suggesstions.
\end{acknowledgements}

\bibliographystyle{aa} % style aa.bst
\bibliography{Halo_Bfield}     % your references Yourfile.bib

\begin{thebibliography}{18}
\expandafter\ifx\csname natexlab\endcsname\relax\def\natexlab#1{#1}\fi

\bibitem[{{Brentjens} \& {de Bruyn}(2005)}]{Brentjens2005}
{Brentjens}, M.~A. \& {de Bruyn}, A.~G. 2005, \aap, 441, 1217

\bibitem[{{Christodoulou} {et~al.}(2016){Christodoulou}, {Gabuzda}, {Knuettel},
  {Contopoulos}, {Kazanas}, \& {Coughlan}}]{Christodoulou2016}
{Christodoulou}, D.~M., {Gabuzda}, D.~C., {Knuettel}, S., {et~al.} 2016, \aap,
  591, A61

\bibitem[{{Contopoulos} {et~al.}(2015){Contopoulos}, {Gabuzda}, \&
  {Kylafis}}]{Contopoulos2015}
{Contopoulos}, I., {Gabuzda}, D., \& {Kylafis}, N. 2015, {The Formation and
  Disruption of Black Hole Jets}, Vol. 414

\bibitem[{{Contopoulos} \& {Kazanas}(1998)}]{Contopoulos1998}
{Contopoulos}, I. \& {Kazanas}, D. 1998, \apj, 508, 859

\bibitem[{{Cordes} \& {Lazio}(2002)}]{Cordes2002}
{Cordes}, J.~M. \& {Lazio}, T.~J.~W. 2002, arXiv e-prints, astro

\bibitem[{{Feain} {et~al.}(2009){Feain}, {Ekers}, {Murphy}, {Gaensler},
  {Macquart}, {Norris}, {Cornwell}, {Johnston-Hollitt}, {Ott}, \&
  {Middelberg}}]{Feain2009}
{Feain}, I.~J., {Ekers}, R.~D., {Murphy}, T., {et~al.} 2009, \apj, 707, 114

\bibitem[{{Gabuzda} {et~al.}(2012){Gabuzda}, {Christodoulou}, {Contopoulos}, \&
  {Kazanas}}]{Gabuzda2012}
{Gabuzda}, D.~C., {Christodoulou}, D.~M., {Contopoulos}, I., \& {Kazanas}, D.
  2012, in Journal of Physics Conference Series, Vol. 355, Journal of Physics
  Conference Series, 012019

\bibitem[{{Irwin} {et~al.}(2012){Irwin}, {Beck}, {Benjamin}, {Dettmar},
  {English}, {Heald}, {Henriksen}, {Johnson}, {Krause}, {Li}, {Miskolczi},
  {Mora}, {Murphy}, {Oosterloo}, {Porter}, {Rand }, {Saikia}, {Schmidt},
  {Strong}, {Walterbos}, {Wang}, \& {Wiegert}}]{Irwin2012}
{Irwin}, J., {Beck}, R., {Benjamin}, R.~A., {et~al.} 2012, \aj, 144, 43

\bibitem[{{Irwin} {et~al.}(2018){Irwin}, {Henriksen}, {We{\.Z}gowiec},
  {Damas-Segovia}, {Wang}, {Krause}, {Heald}, {Dettmar}, {Li}, {Wiegert},
  {Stein}, {Braun}, {Im}, {Schmidt}, {Macdonald}, {Miskolczi}, {Merritt},
  {Mora-Partiarroyo}, {Saikia}, {Sotomayor}, \& {Yang}}]{Irwin2018}
{Irwin}, J.~A., {Henriksen}, R.~N., {We{\.Z}gowiec}, M., {et~al.} 2018, \mnras,
  476, 5057

\bibitem[{{K{\"o}nigl}(2010)}]{Konigl2010}
{K{\"o}nigl}, A. 2010, \mnras, 407, L79

\bibitem[{{Krause} {et~al.}(2020){Krause}, {Irwin}, {Schmidt}, {Stein},
  {Miskolczi}, {Carolina Mora-Partiarroyo}, {Wiegert}, {Beck}, {Stil}, {Heald},
  {Li}, {Damas-Segovia}, {Vargas}, {Rand}, {West}, {Walterbos}, {Dettmar},
  {English}, \& {Woodfinden}}]{Krause2020}
{Krause}, M., {Irwin}, J., {Schmidt}, P., {et~al.} 2020, \aap, 639, A112

\bibitem[{{Kronberg}(2016)}]{Kronberg2016}
{Kronberg}, P.~P. 2016, {Cosmic Magnetic Fields}

\bibitem[{{Lynden-Bell}(2013)}]{Lynden-Bell2013}
{Lynden-Bell}, D. 2013, The Observatory, 133, 266

\bibitem[{{Makarov} {et~al.}(2014){Makarov}, {Prugniel}, {Terekhova},
  {Courtois}, \& {Vauglin}}]{Makarov2014}
{Makarov}, D., {Prugniel}, P., {Terekhova}, N., {Courtois}, H., \& {Vauglin},
  I. 2014, \aap, 570, A13

\bibitem[{{Mora-Partiarroyo} {et~al.}(2019){Mora-Partiarroyo}, {Krause},
  {Basu}, {Beck}, {Wiegert}, {Irwin}, {Henriksen}, {Stein}, {Vargas}, {Heesen},
  {Walterbos}, {Rand}, {Heald}, {Li}, {Kamieneski}, \&
  {English}}]{Mora-Partiarroyo2019}
{Mora-Partiarroyo}, S.~C., {Krause}, M., {Basu}, A., {et~al.} 2019, \aap, 632,
  A11

\bibitem[{{Oppermann} {et~al.}(2015){Oppermann}, {Junklewitz}, {Greiner},
  {En{\ss}lin}, {Akahori}, {Carretti}, {Gaensler}, {Goobar}, {Harvey-Smith},
  {Johnston-Hollitt}, {Pratley}, {Schnitzeler}, {Stil}, \&
  {Vacca}}]{Oppermann2015}
{Oppermann}, N., {Junklewitz}, H., {Greiner}, M., {et~al.} 2015, \aap, 575,
  A118

\bibitem[{{Pshirkov} {et~al.}(2011){Pshirkov}, {Tinyakov}, {Kronberg}, \&
  {Newton-McGee}}]{Pshirkov2011}
{Pshirkov}, M.~S., {Tinyakov}, P.~G., {Kronberg}, P.~P., \& {Newton-McGee},
  K.~J. 2011, \apj, 738, 192

\bibitem[{{Wiegert} {et~al.}(2015){Wiegert}, {Irwin}, {Miskolczi}, {Schmidt},
  {Mora}, {Damas-Segovia}, {Stein}, {English}, {Rand}, {Santistevan},
  {Walterbos}, {Krause}, {Beck}, {Dettmar}, {Kepley}, {Wezgowiec}, {Wang},
  {Heald}, {Li}, {MacGregor}, {Johnson}, {Strong}, {DeSouza}, \&
  {Porter}}]{Wiegert2015}
{Wiegert}, T., {Irwin}, J., {Miskolczi}, A., {et~al.} 2015, \aj, 150, 81

\end{thebibliography}

\begin{appendix}

\section{RM maps}
\label{app:RMmaps}

RM maps for the 24 galaxies in the CHANG-ES sample that contribute data within the white ellipse of Fig.~\ref{fig:main}. The name of each galaxy appears on the top left within the corresponding panel. The RM maps were created following the methodology described in Sect.~\ref{sec:RM analysis}. The maps contain ten contours of the total intensity (Stokes $I$) at C band, with RM overlaid according to the color wedges. The total intensity contours are distributed on a log scale between the 2$\sigma$ of the r.m.s. and the peak flux density of the total intensity map at C band. We note that the RM range is limited to about $\pm$50~rad/m$^2$ due to the data analysis technique (see text for details). Therefore the RM may be underestimated, especially in regions with high electron density or magnetic field strength.

% -----------------------------------------------------------------------
\begin{figure*}%[!ht] 
\centering
\begin{tabular}{cc}
\includegraphics[trim={0 0 0 0}, clip, width=8cm]{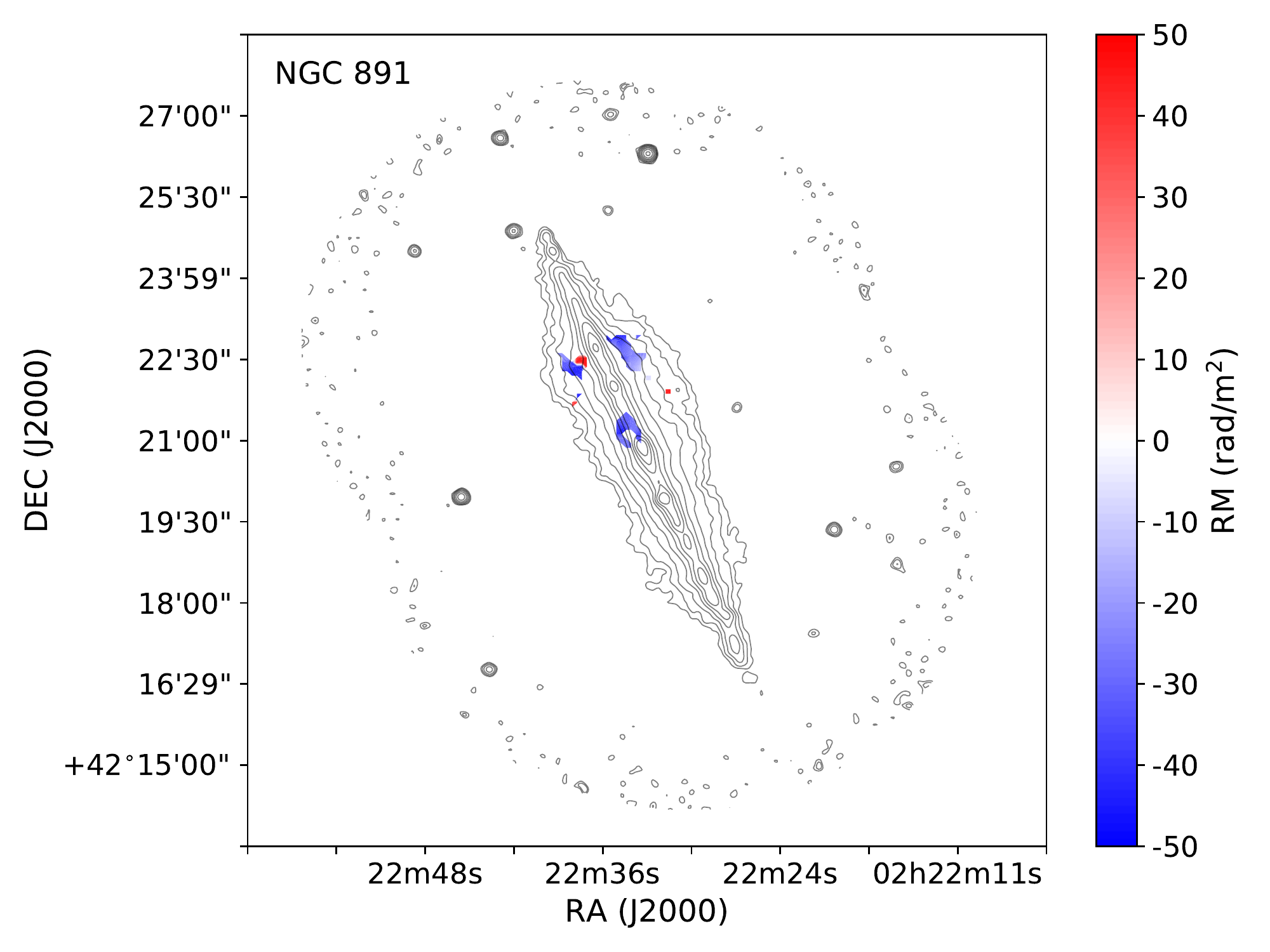}    &
\includegraphics[trim={0 0 0 0}, clip, width=8cm]{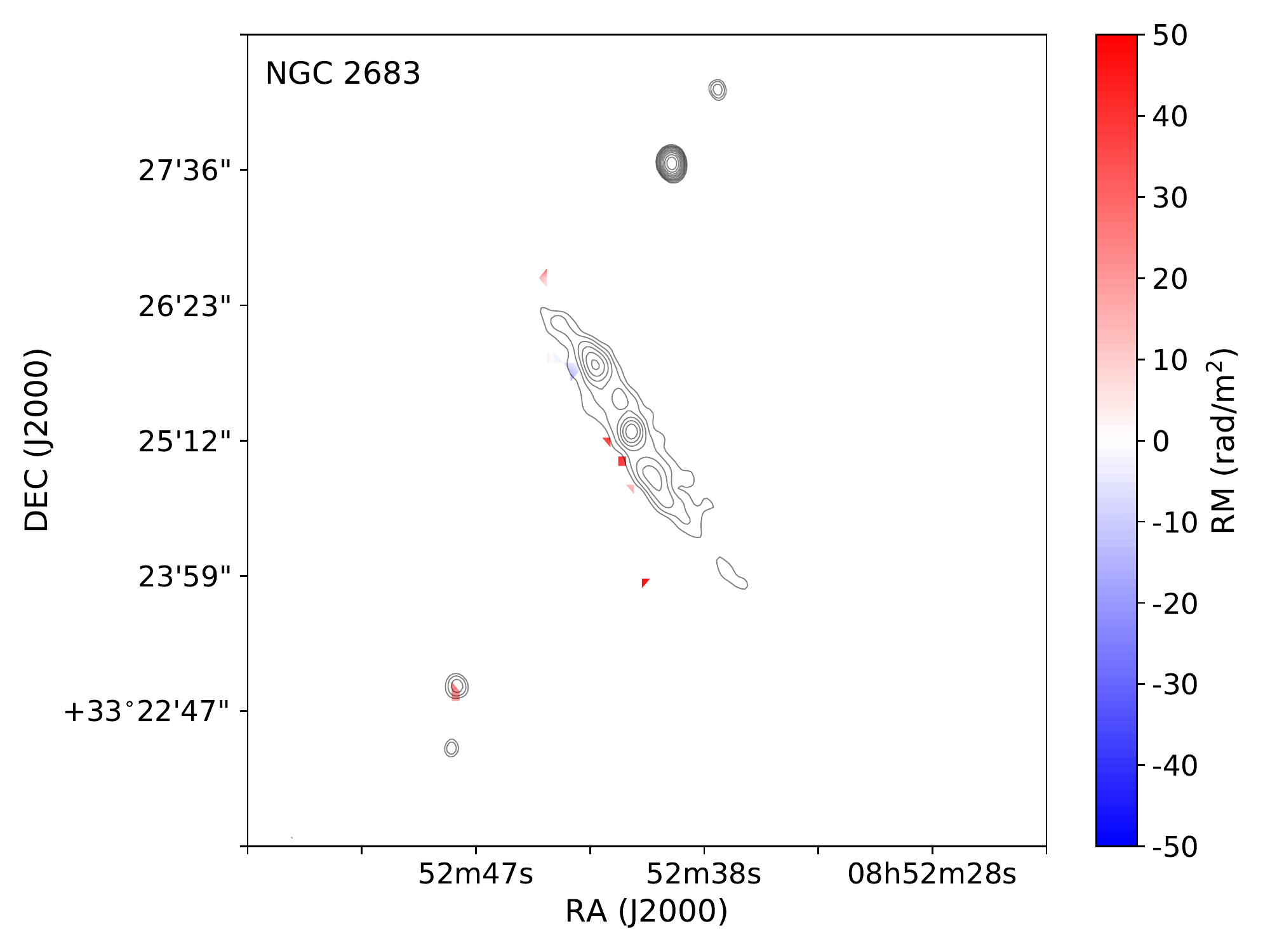}   \\
\includegraphics[trim={0 0 0 0}, clip, width=8cm]{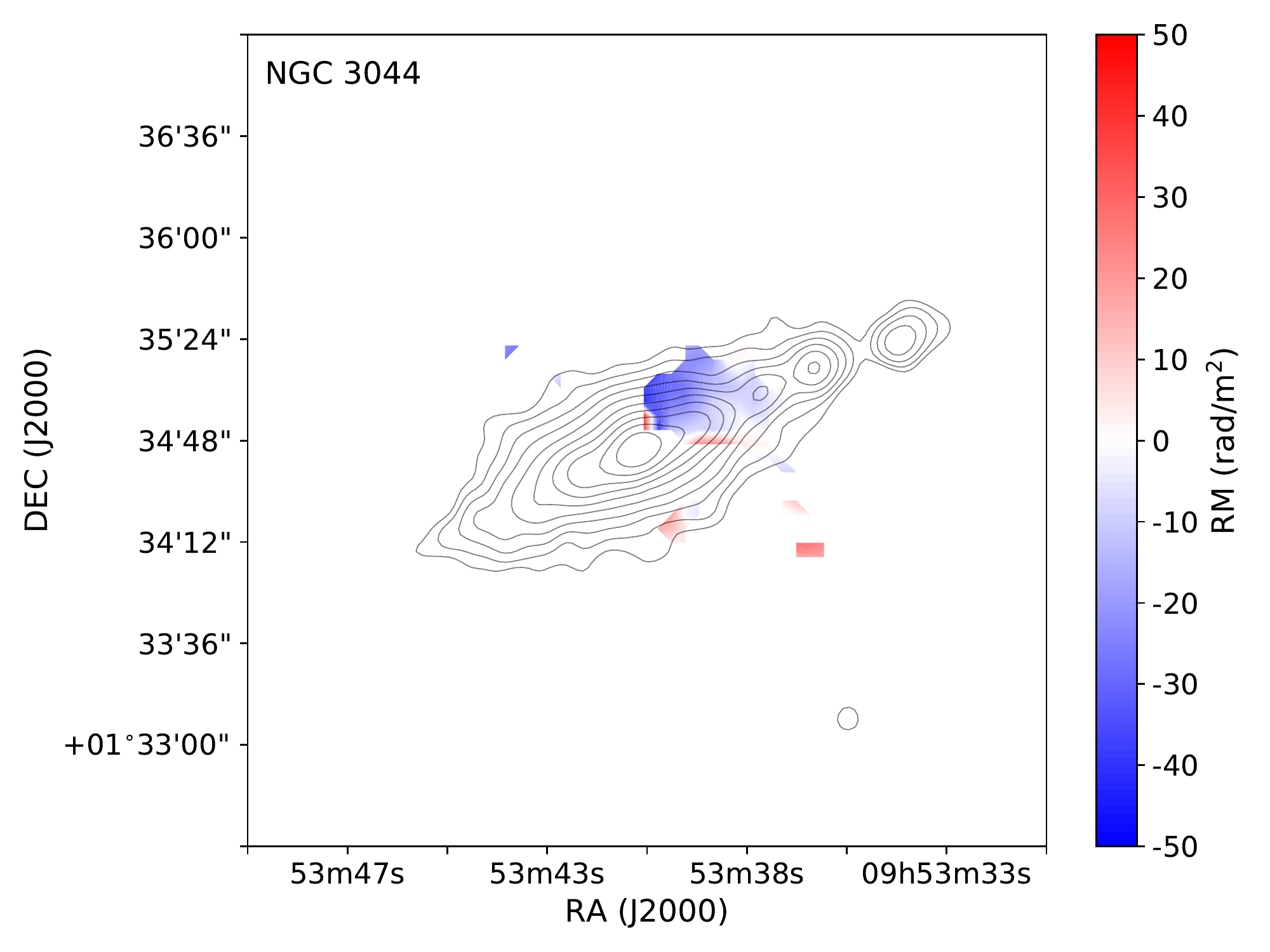}   &
\includegraphics[trim={0 0 0 0}, clip, width=8cm]{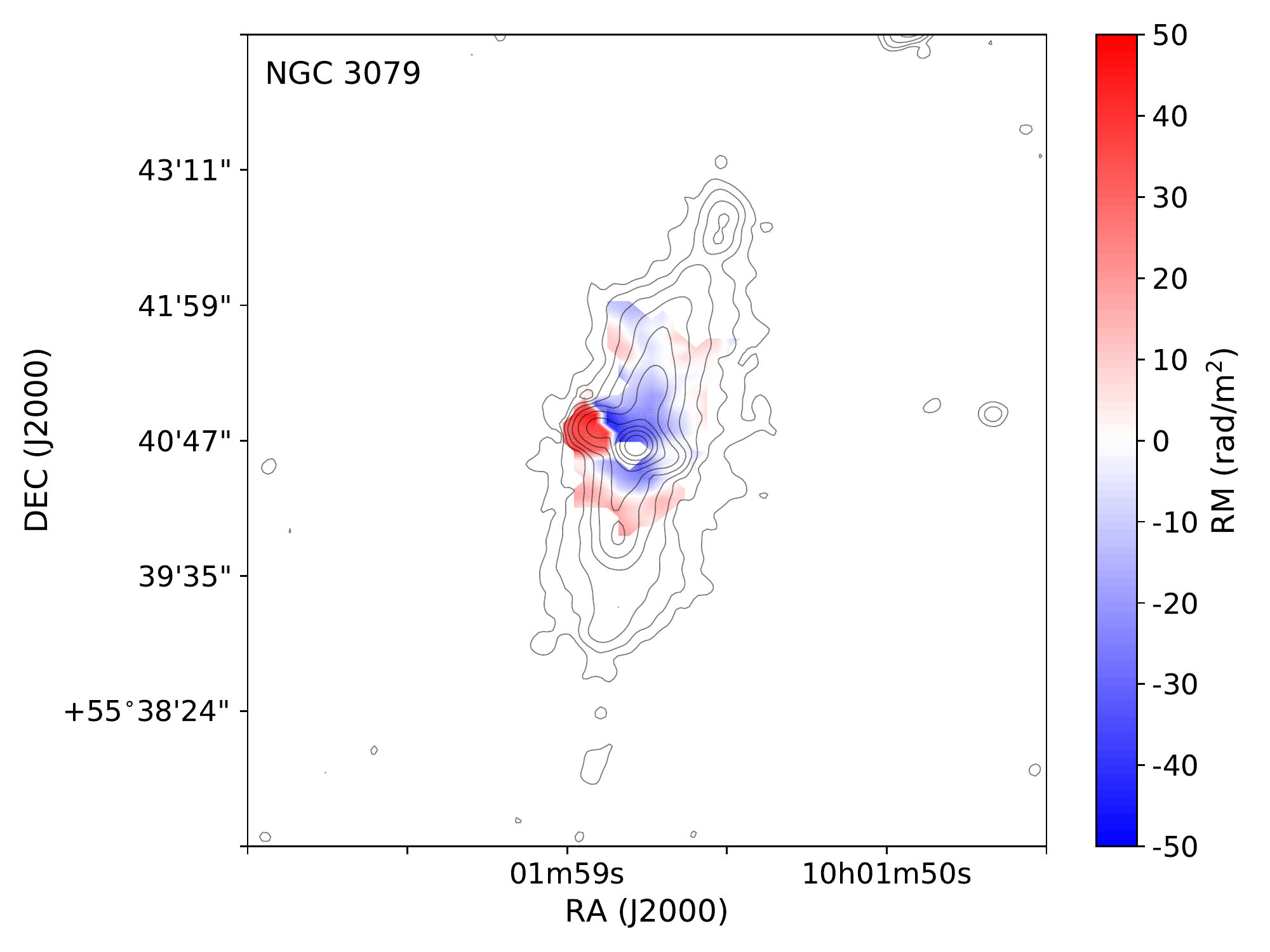}   \\
\includegraphics[trim={0 0 0 0}, clip, width=8cm]{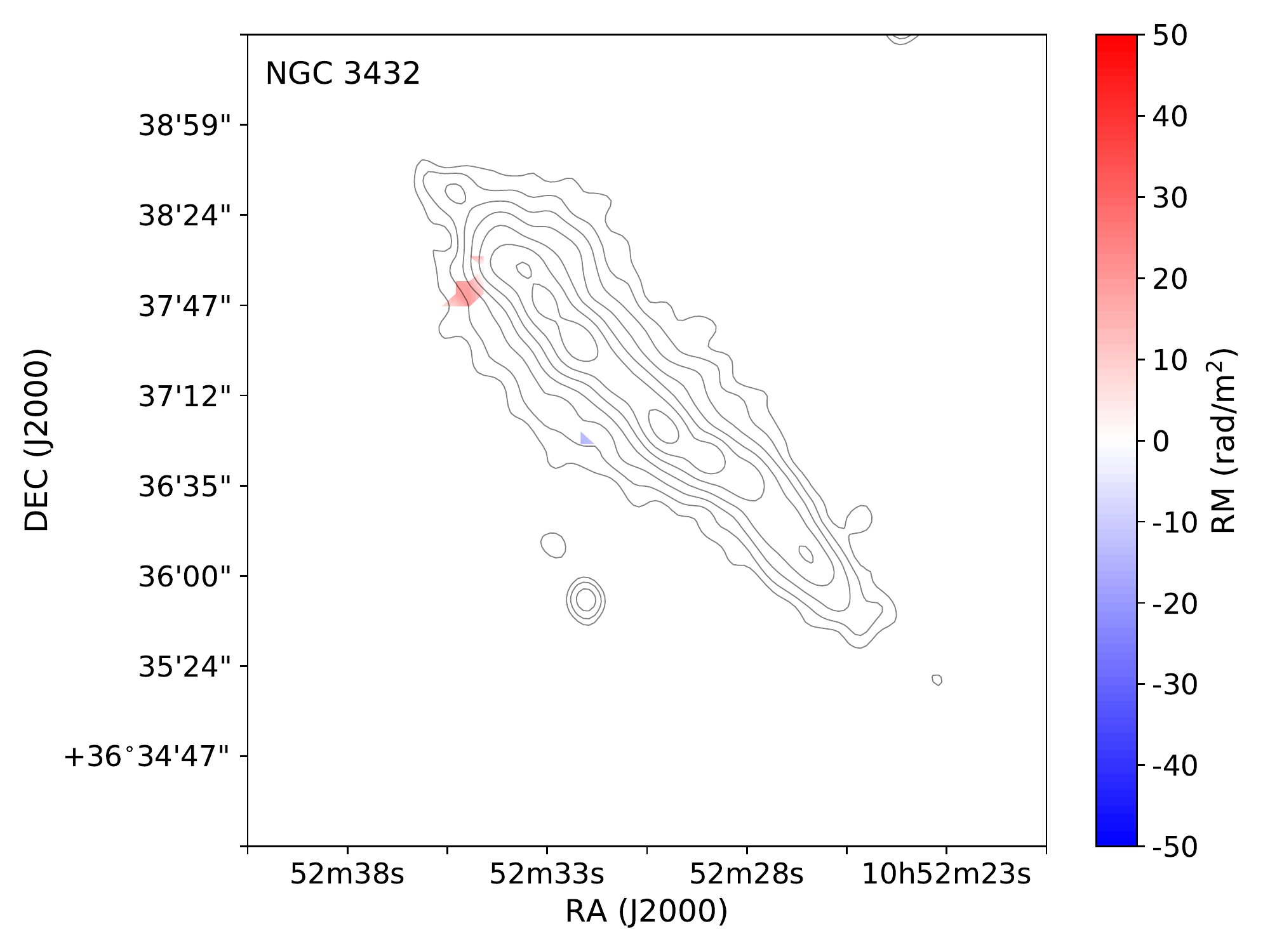}   &
\includegraphics[trim={0 0 0 0}, clip, width=8cm]{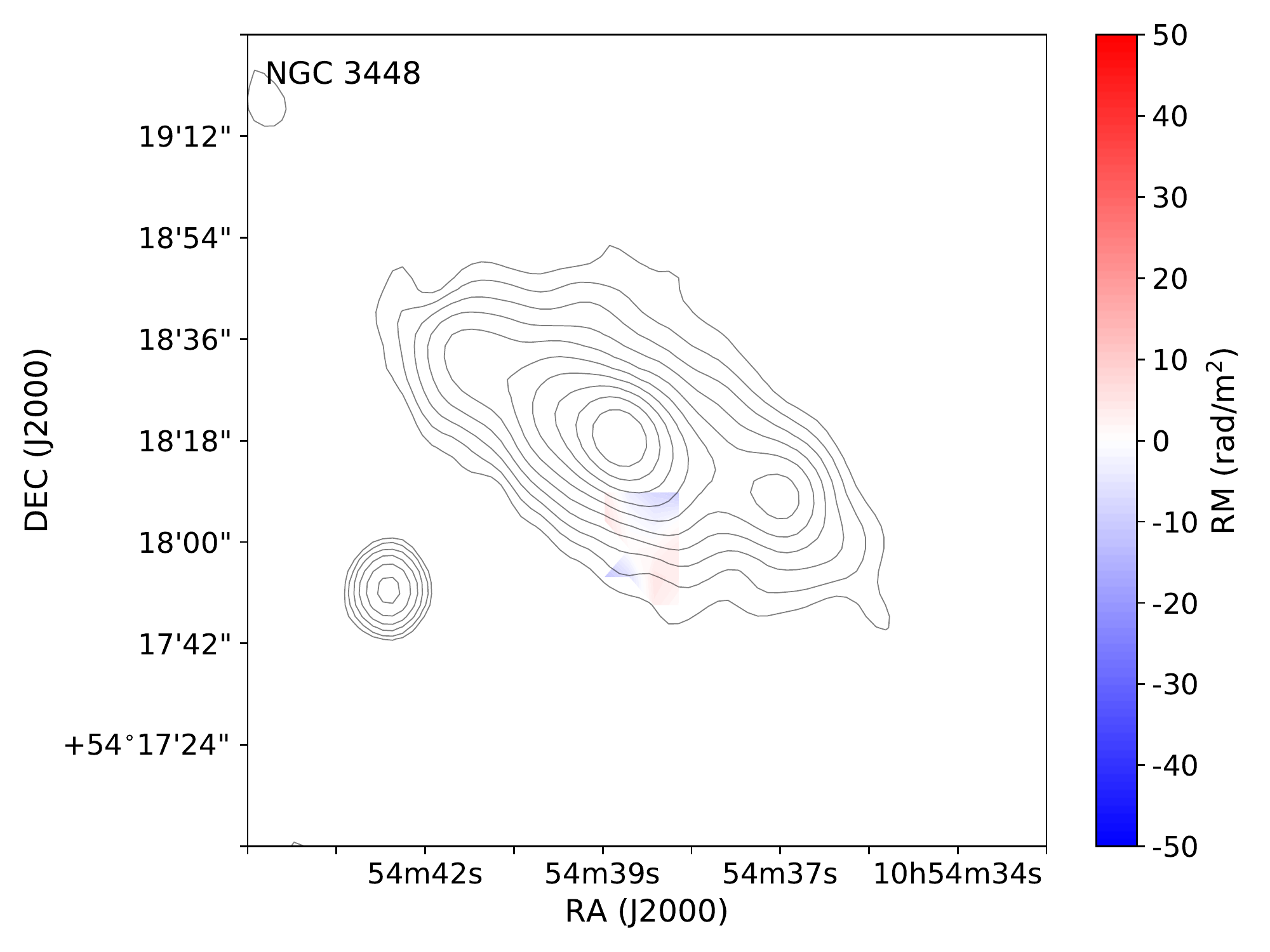}   \\
\includegraphics[trim={0 0 0 0}, clip, width=8cm]{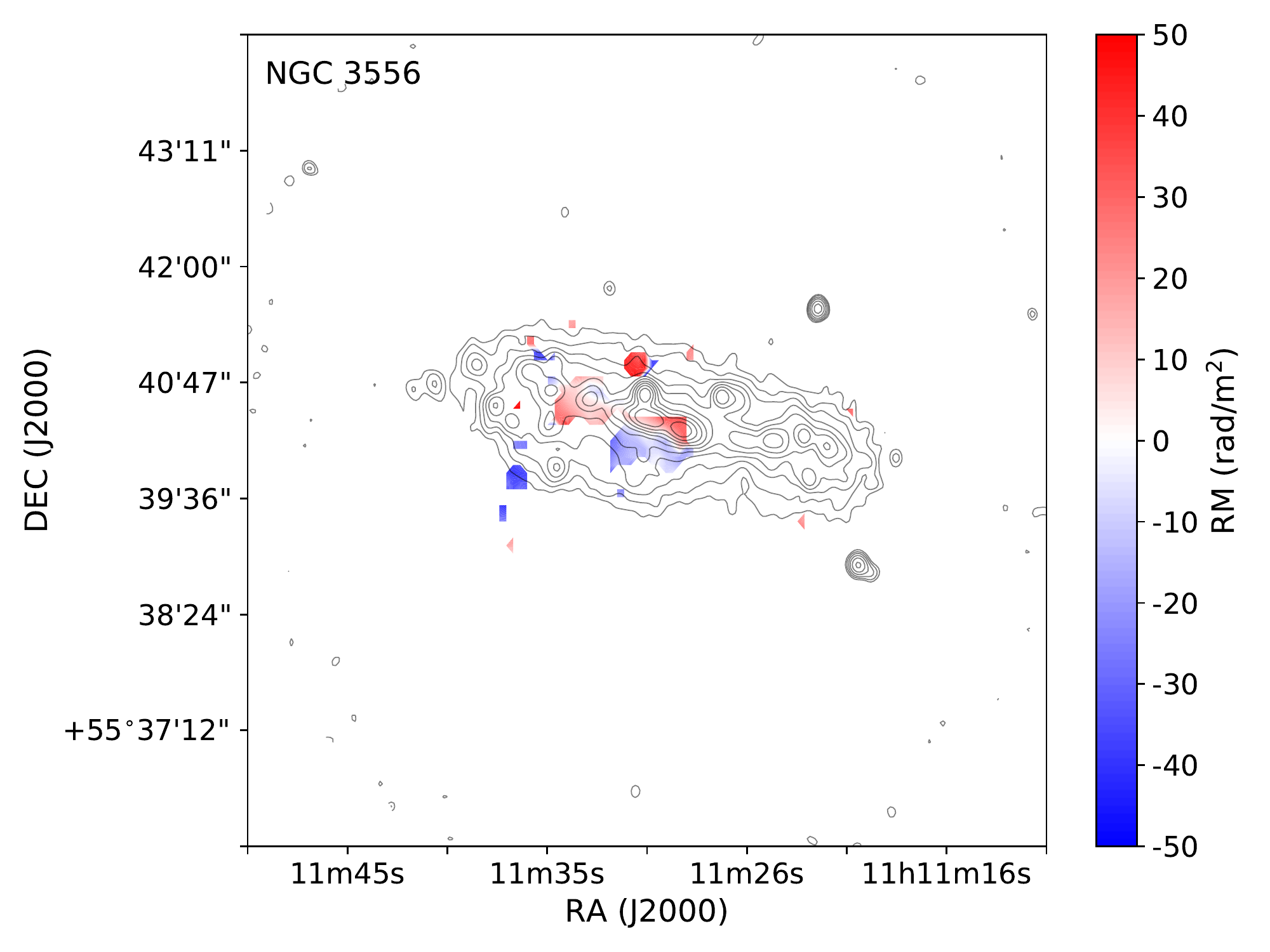}   &
\includegraphics[trim={0 0 0 0}, clip, width=8cm]{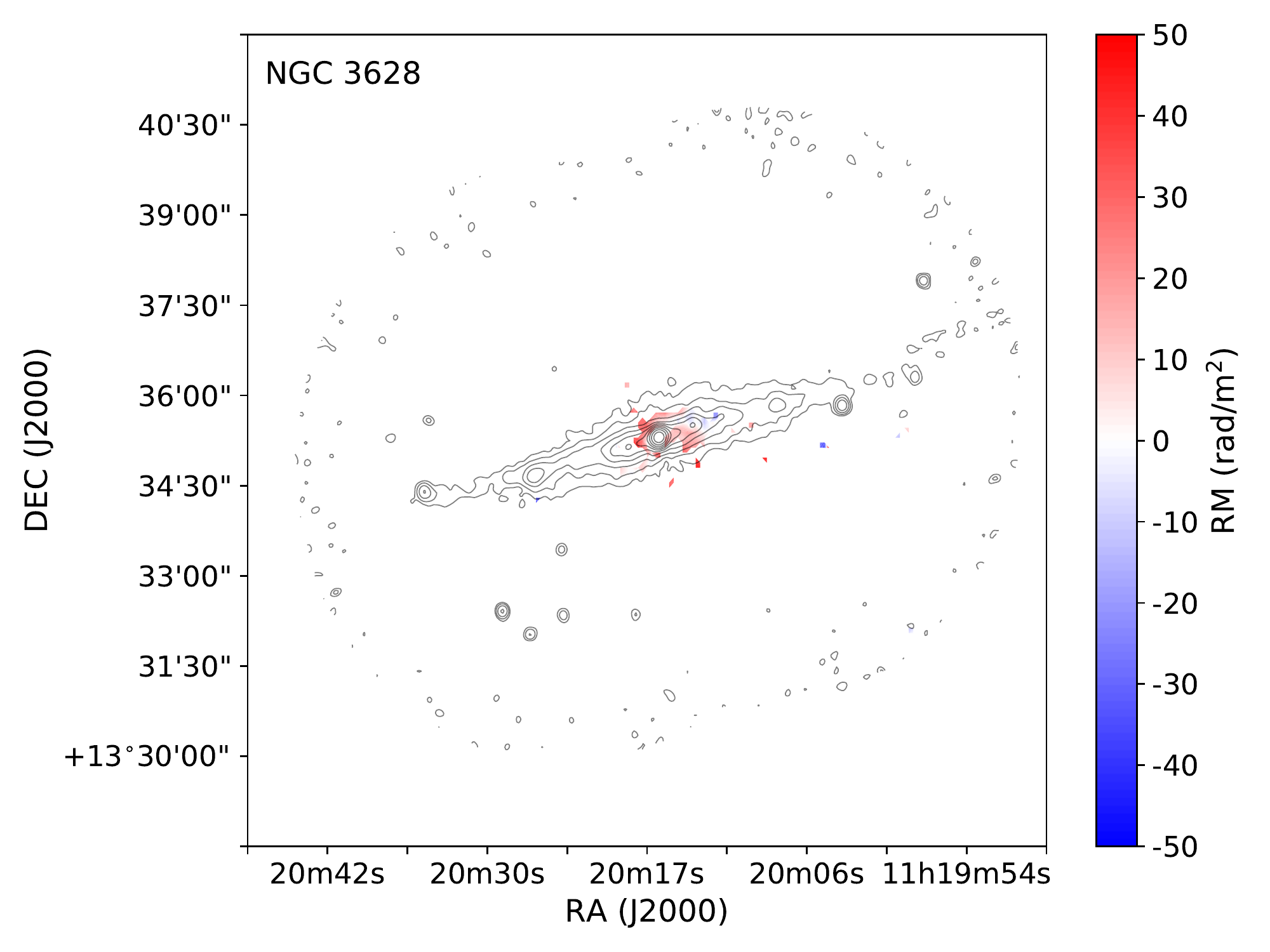}   \\
\end{tabular}
\caption{RM maps }
\label{fig:RMmaps1}
\end{figure*}

\begin{figure*}%[!ht] 
\centering
\begin{tabular}{cc}
\includegraphics[trim={0 0 0 0}, clip, width=8cm]{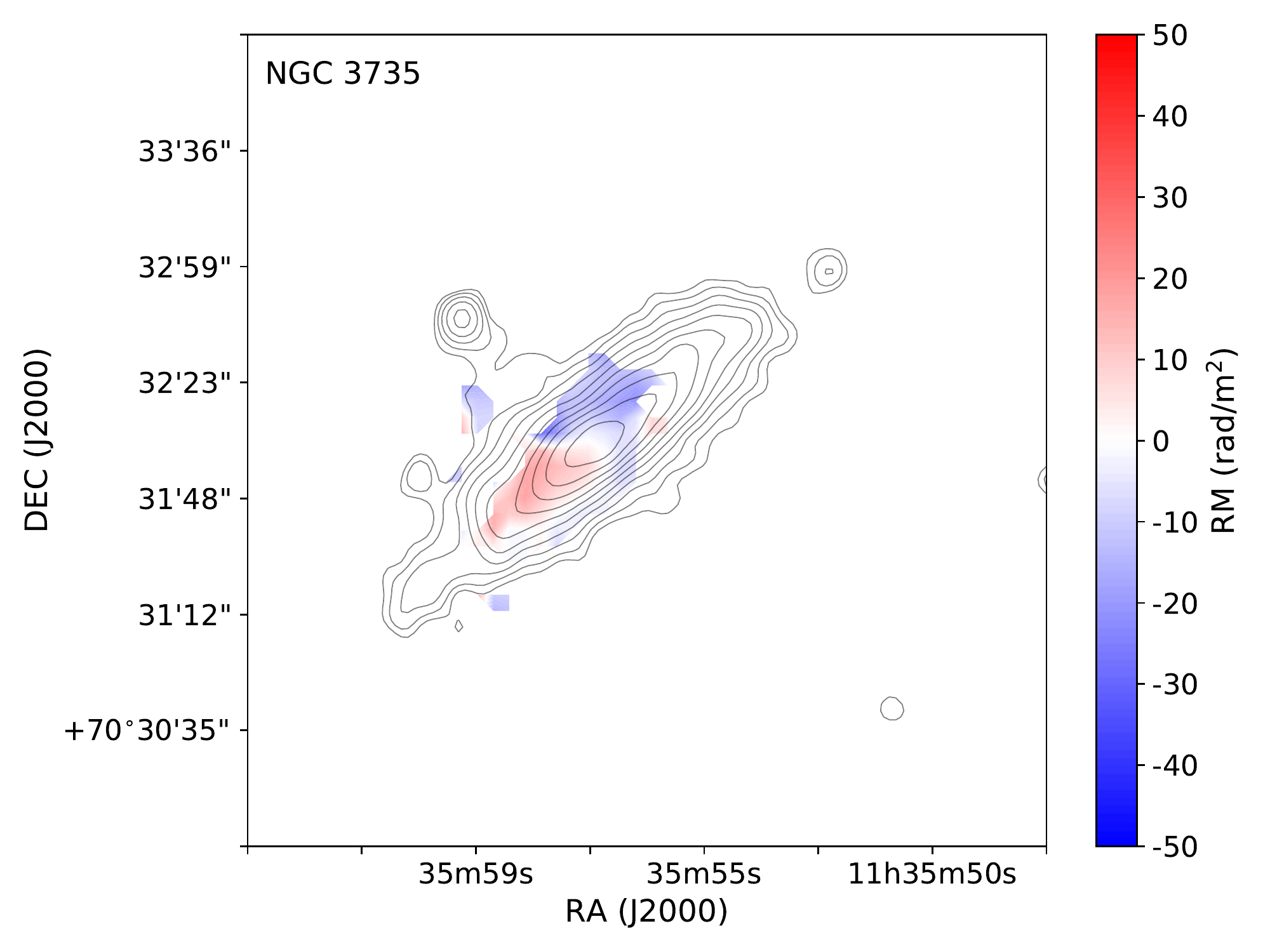}   &
\includegraphics[trim={0 0 0 0}, clip, width=8cm]{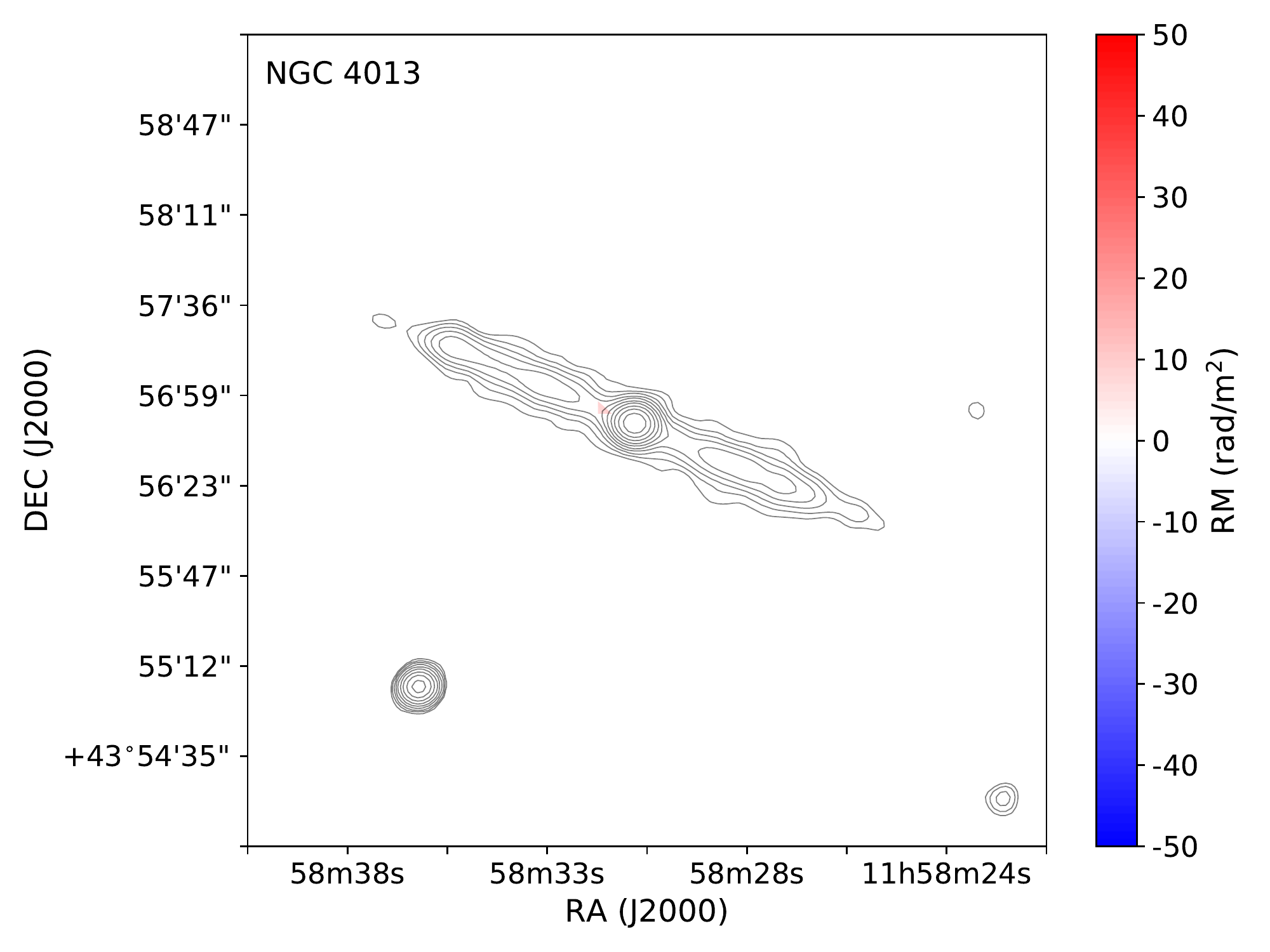}   \\
\includegraphics[trim={0 0 0 0}, clip, width=8cm]{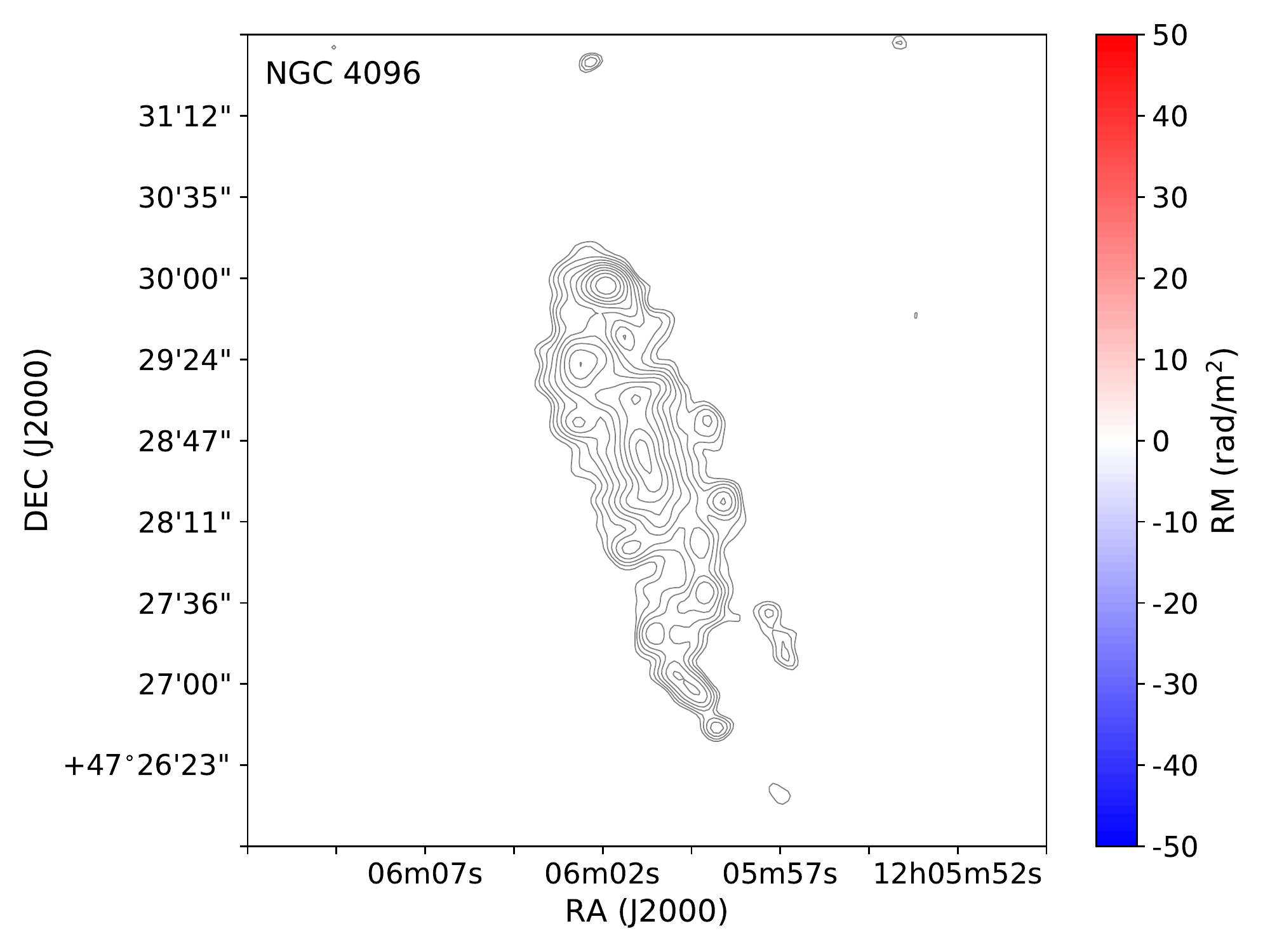}   &
\includegraphics[trim={0 0 0 0}, clip, width=8cm]{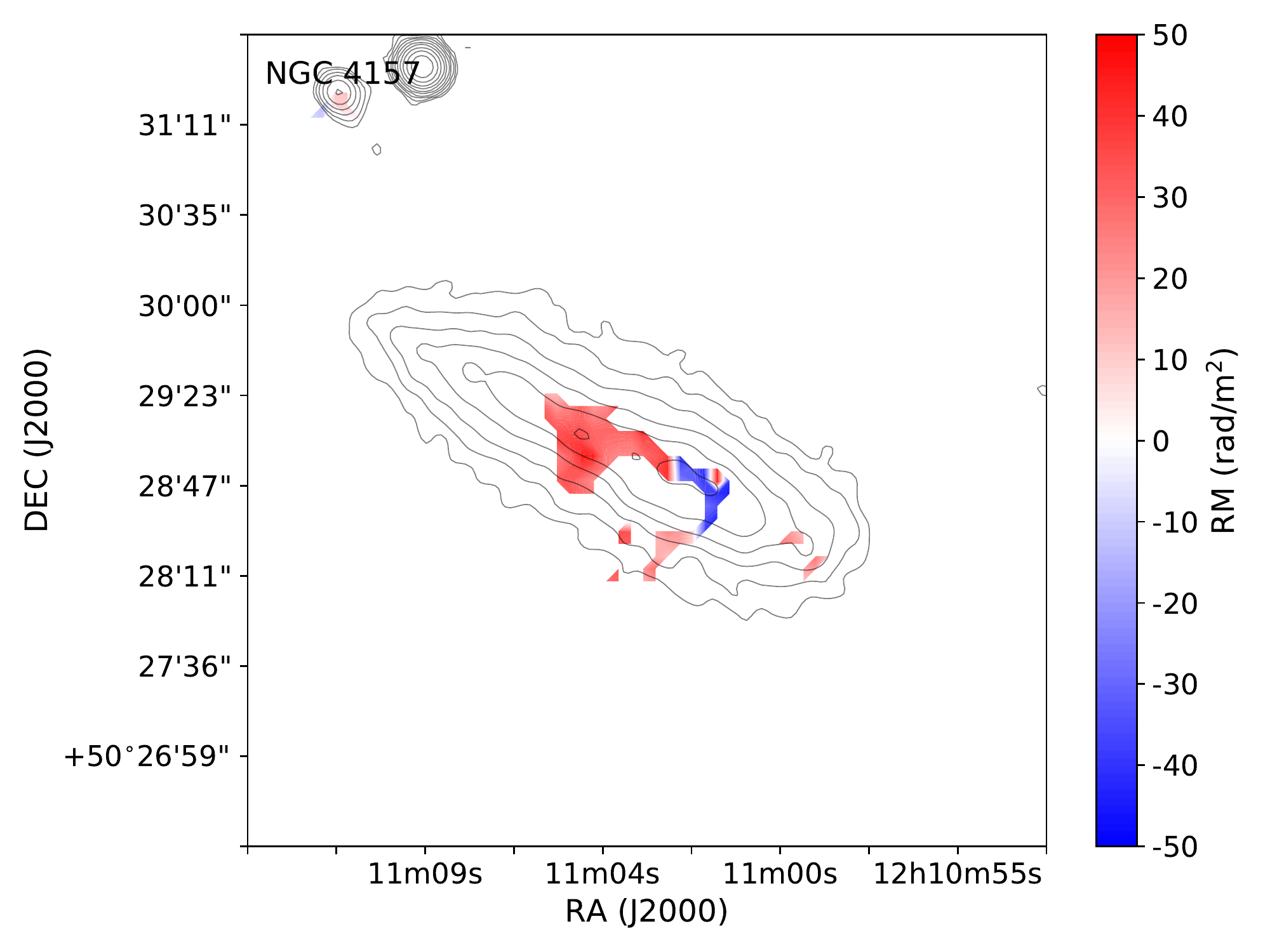}   \\
\includegraphics[trim={0 0 0 0}, clip, width=8cm]{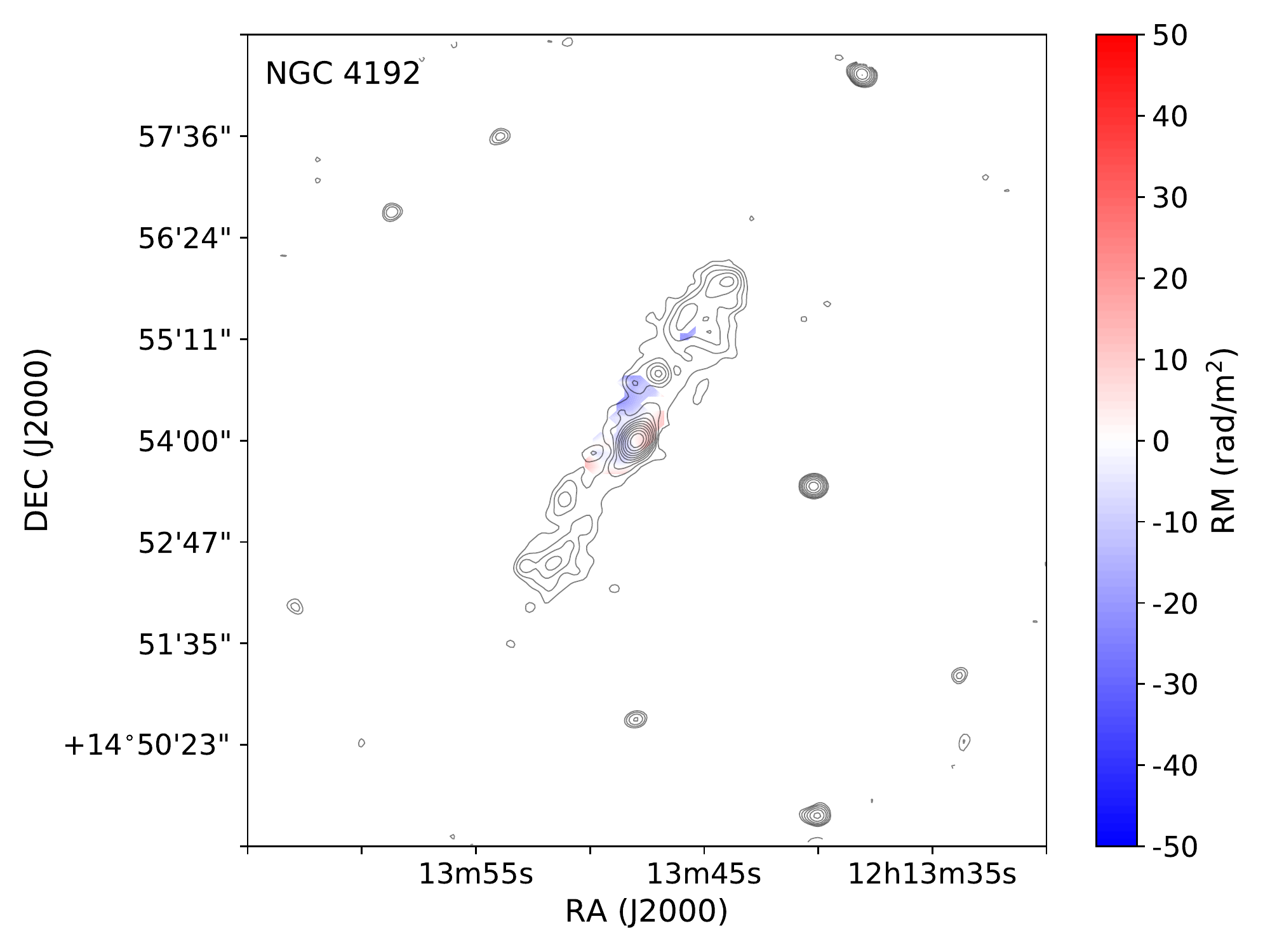}   &
\includegraphics[trim={0 0 0 0}, clip, width=8cm]{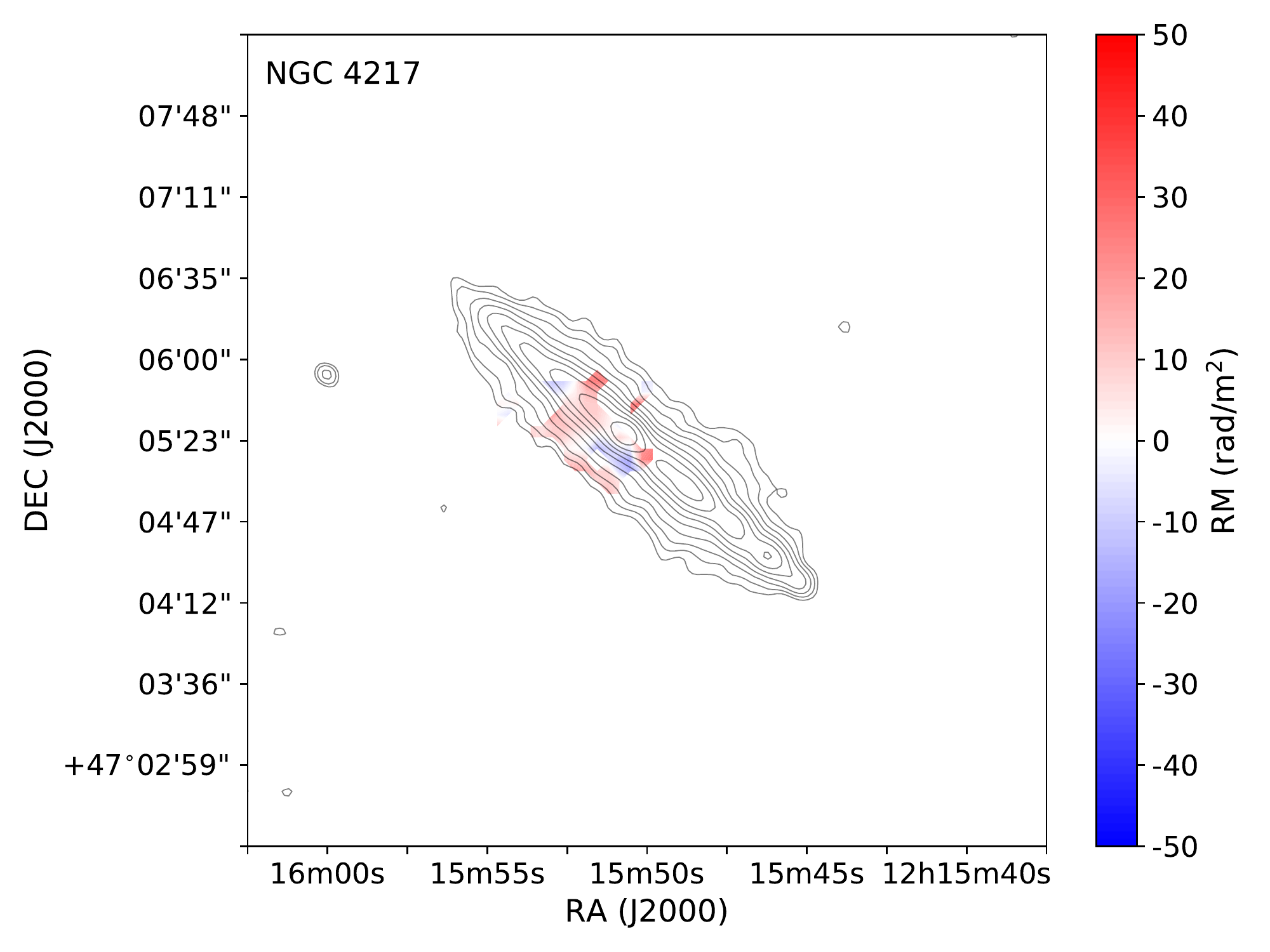}   \\
\includegraphics[trim={0 0 0 0}, clip, width=8cm]{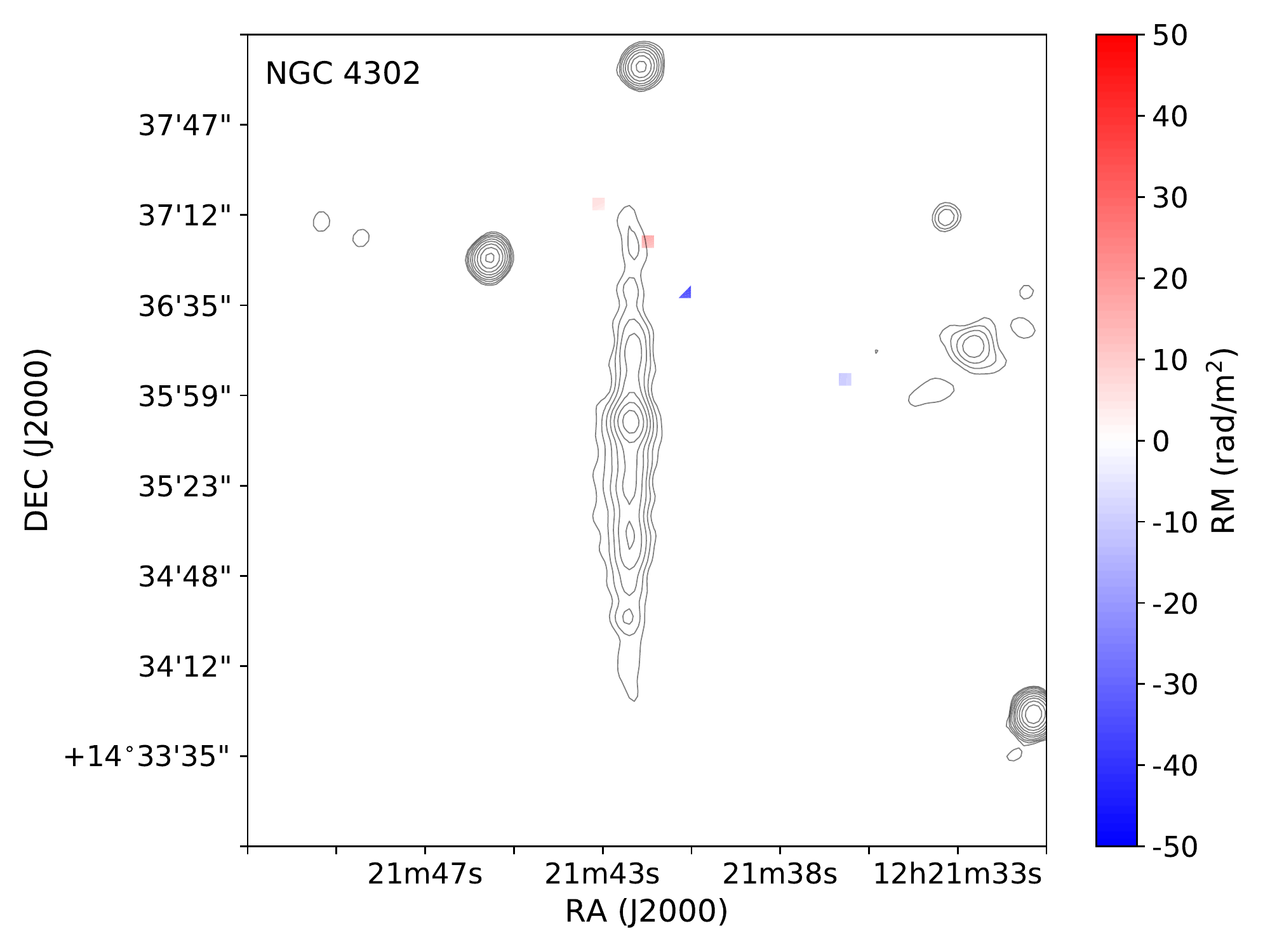}   &
\includegraphics[trim={0 0 0 0}, clip, width=8cm]{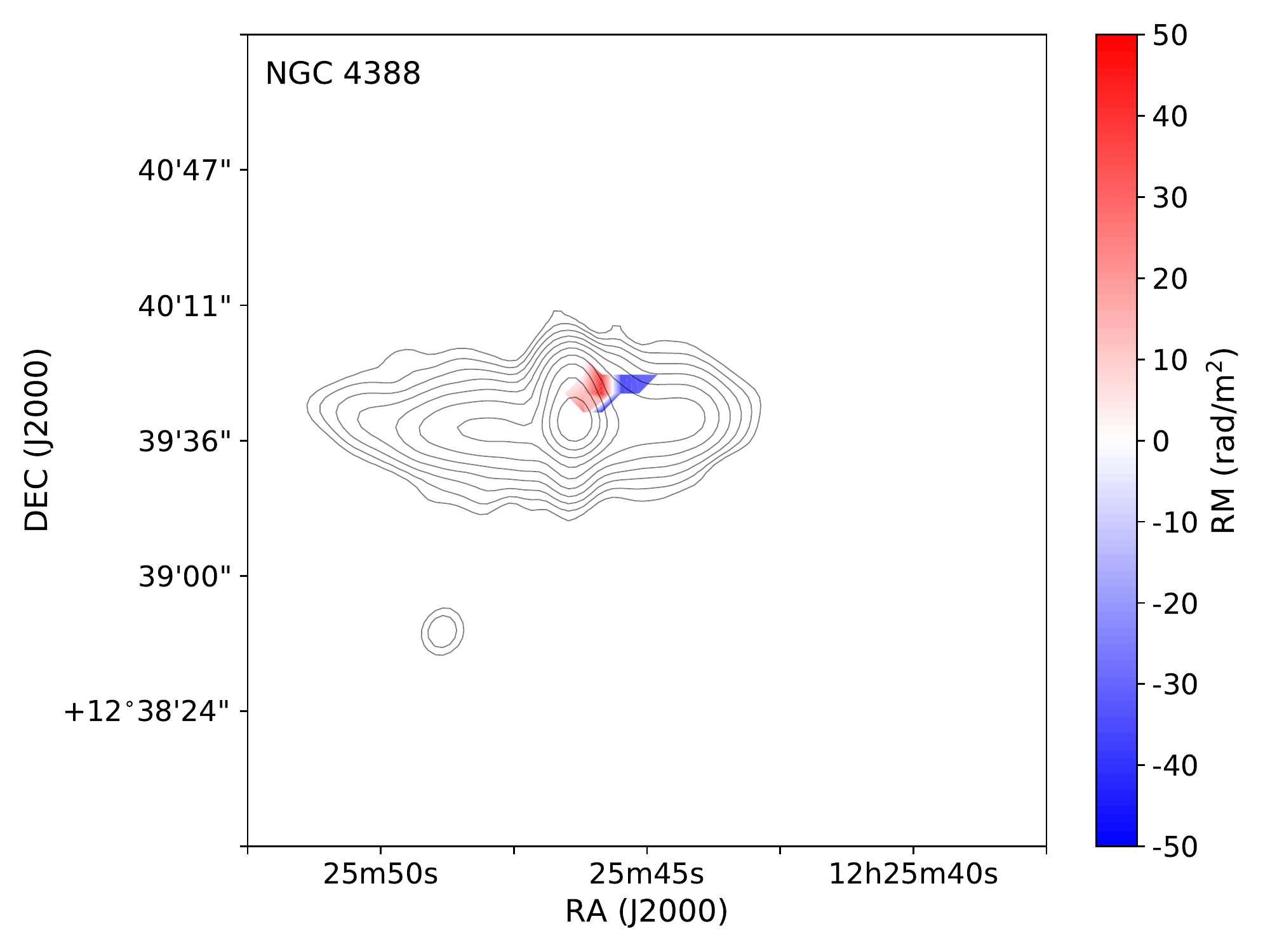}   \\
\end{tabular}
\caption{RM maps (continued)}
\label{fig:RMmaps2}
\end{figure*}

\begin{figure*}%[!ht] 
\centering
\begin{tabular}{cc}
\includegraphics[trim={0 0 0 0}, clip, width=8cm]{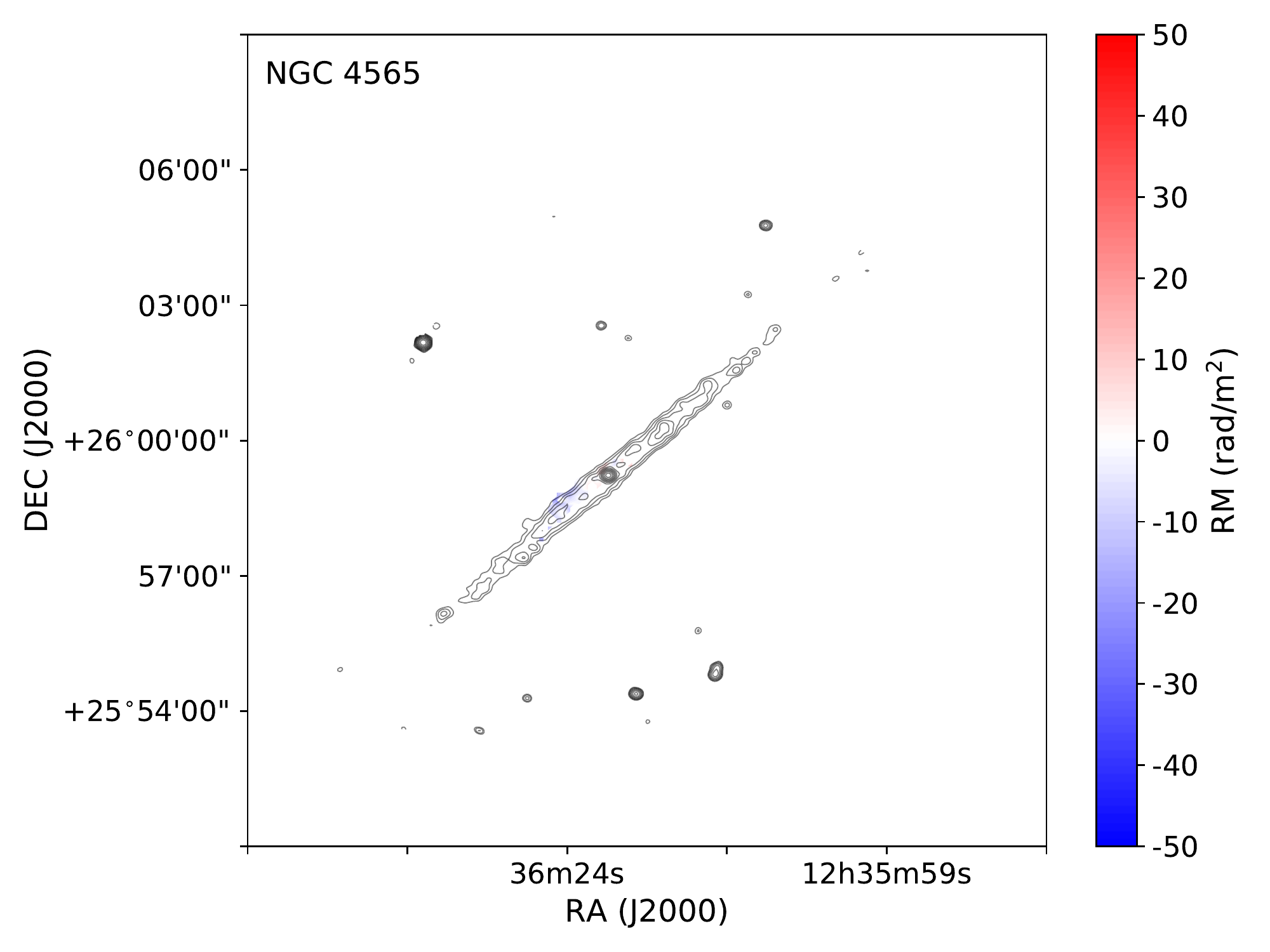}   &
\includegraphics[trim={0 0 0 0}, clip, width=8cm]{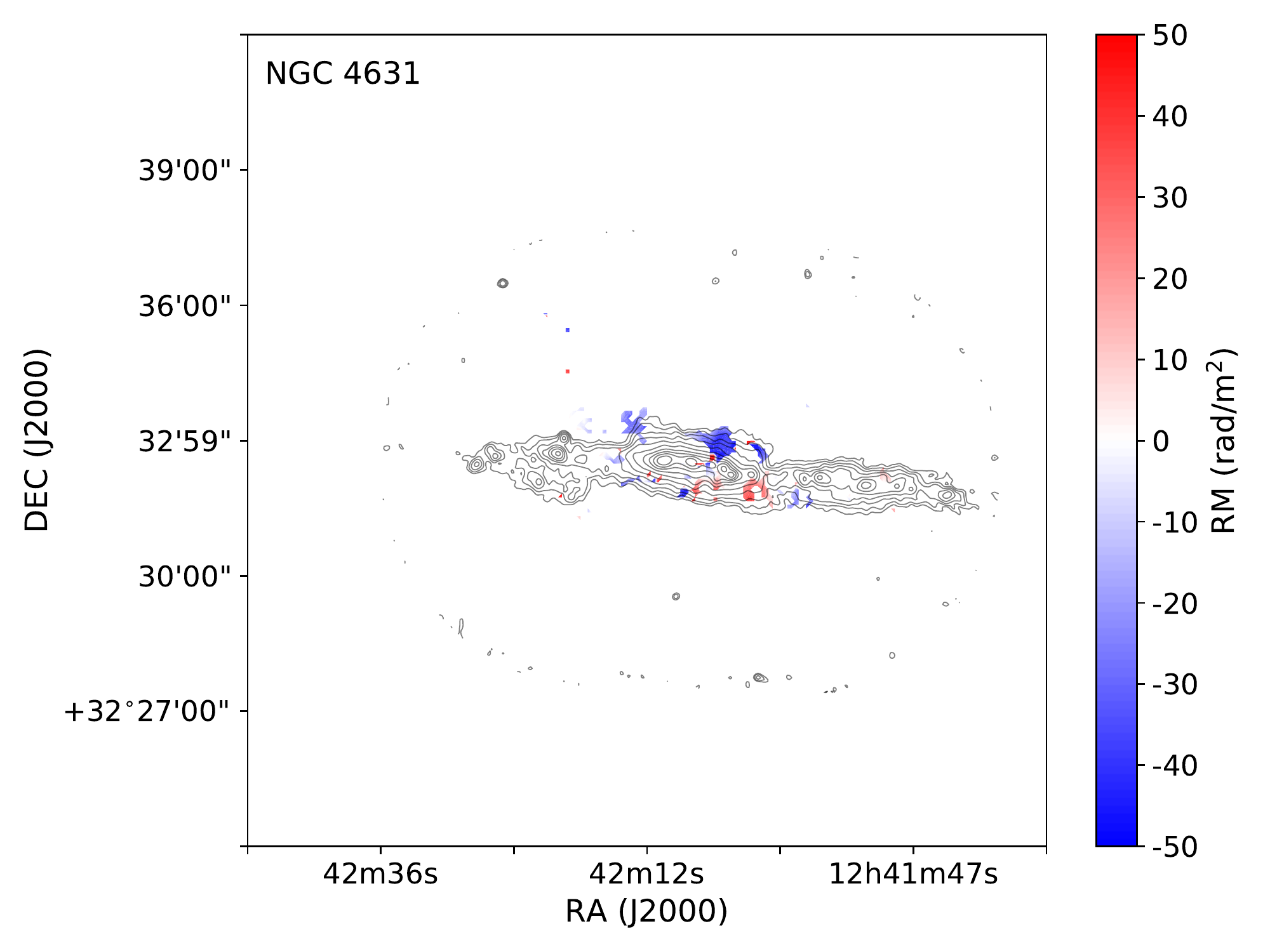}   \\
\includegraphics[trim={0 0 0 0}, clip, width=8cm]{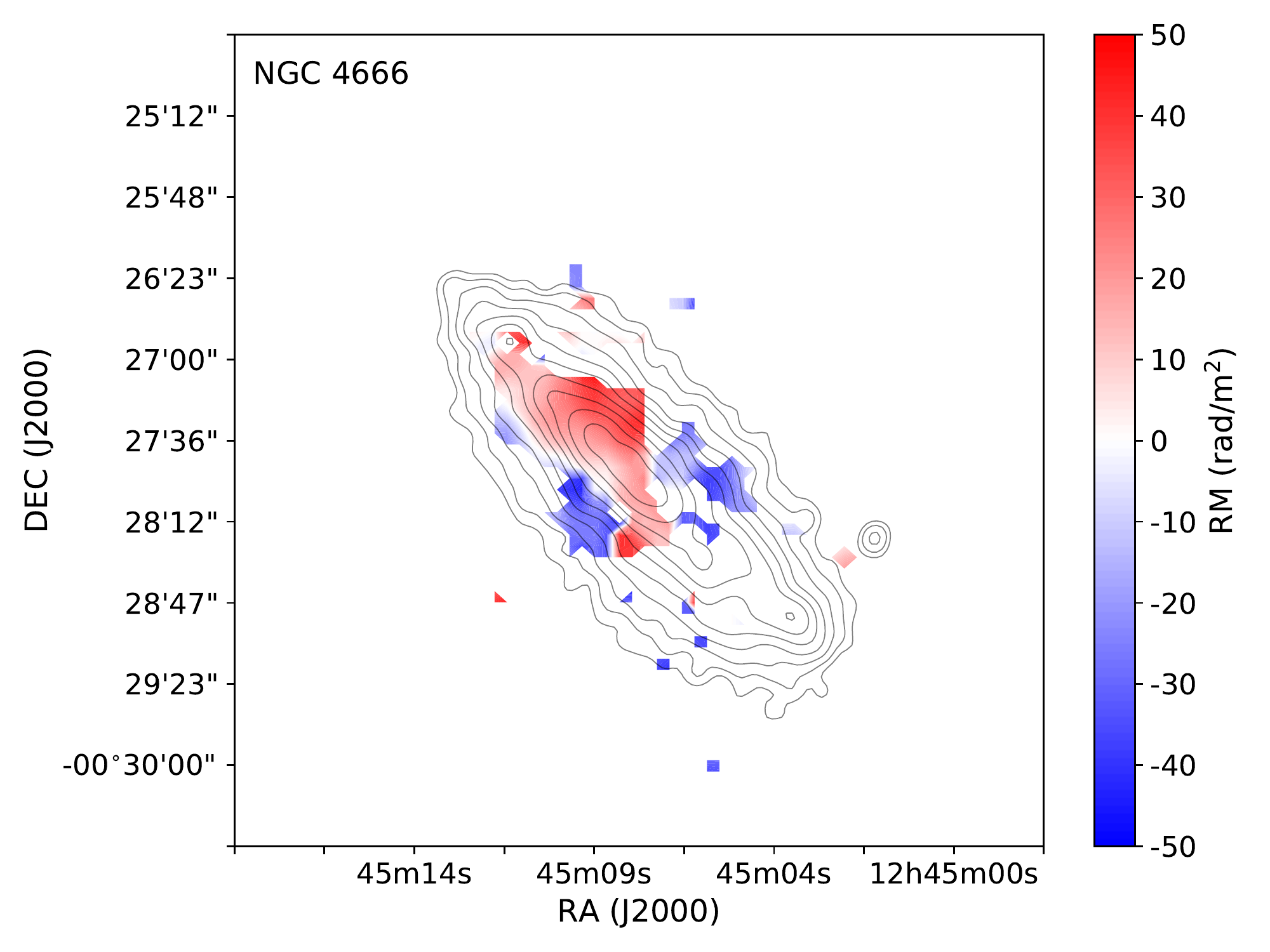}   &
\includegraphics[trim={0 0 0 0}, clip, width=8cm]{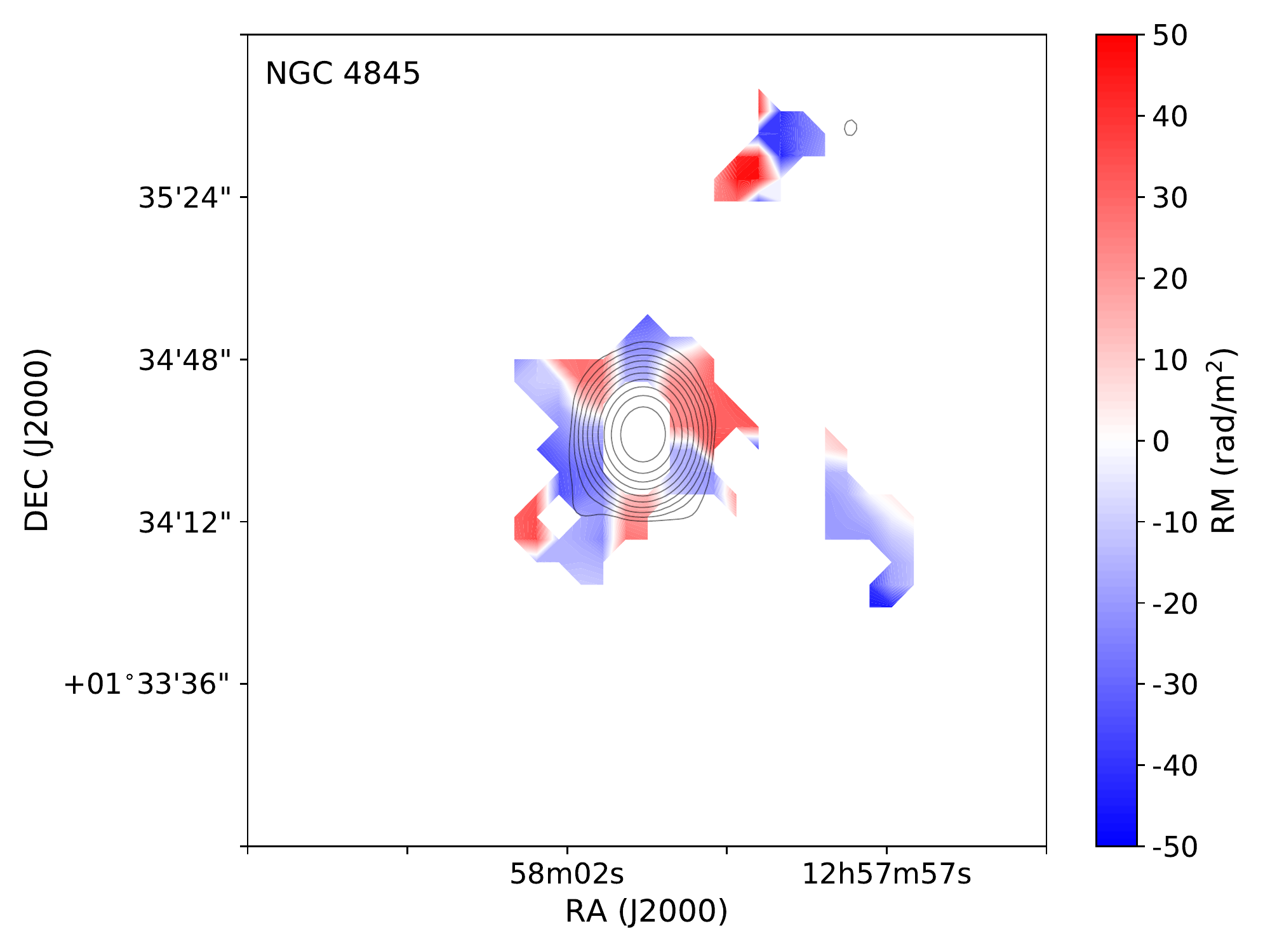}   \\
\includegraphics[trim={0 0 0 0}, clip, width=8cm]{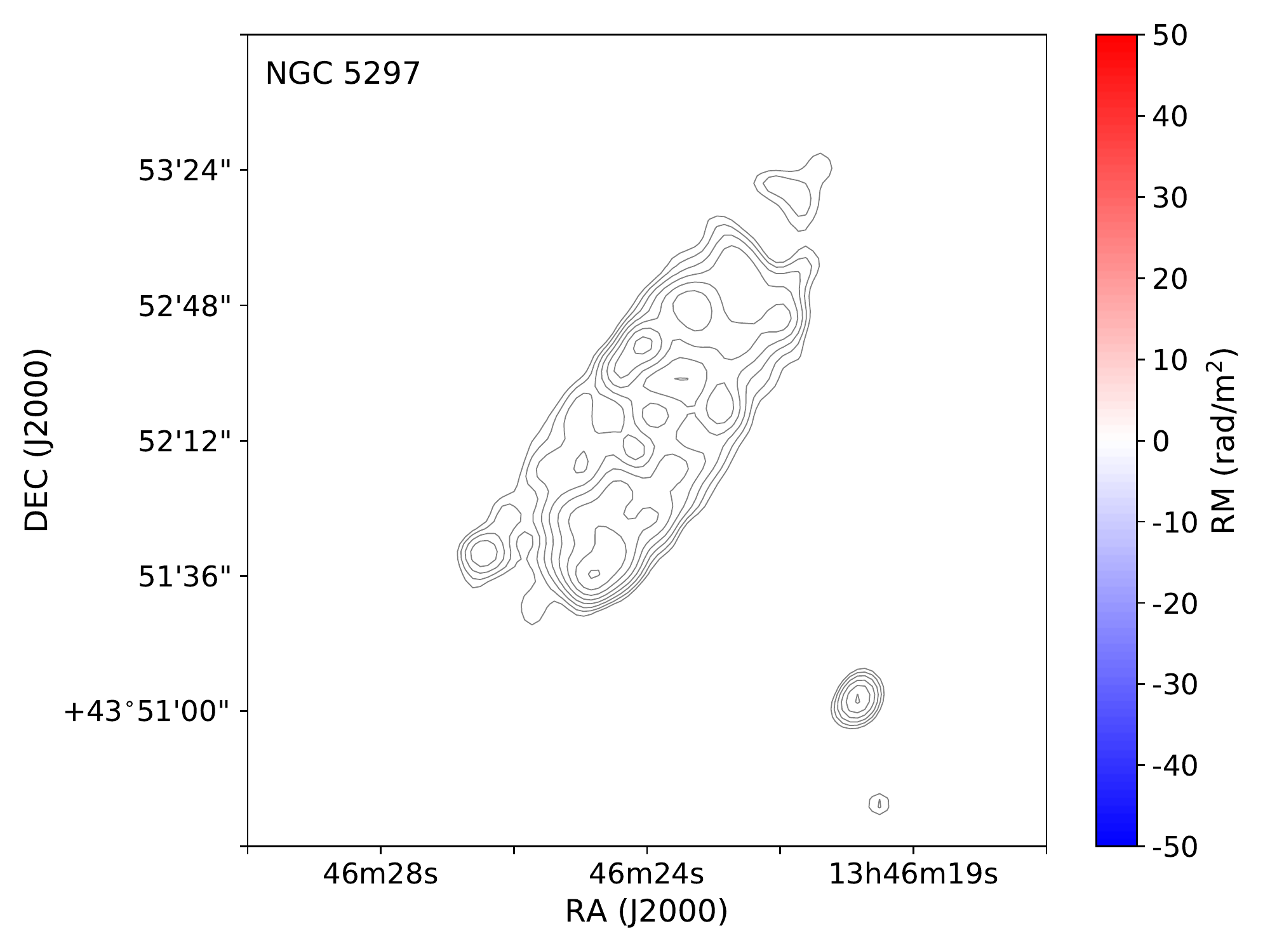}   &
\includegraphics[trim={0 0 0 0}, clip, width=8cm]{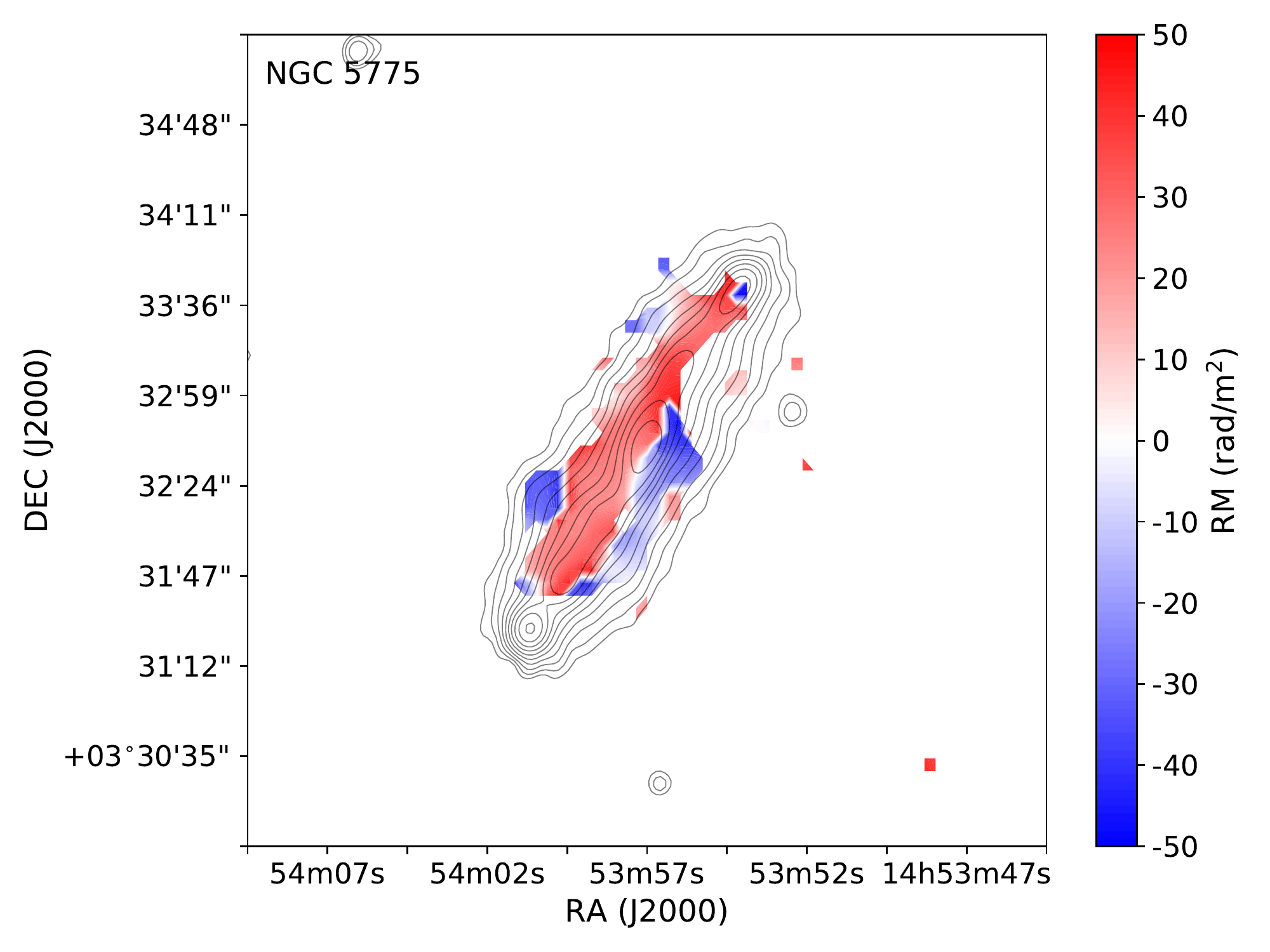}   \\
\includegraphics[trim={0 0 0 0}, clip, width=8cm]{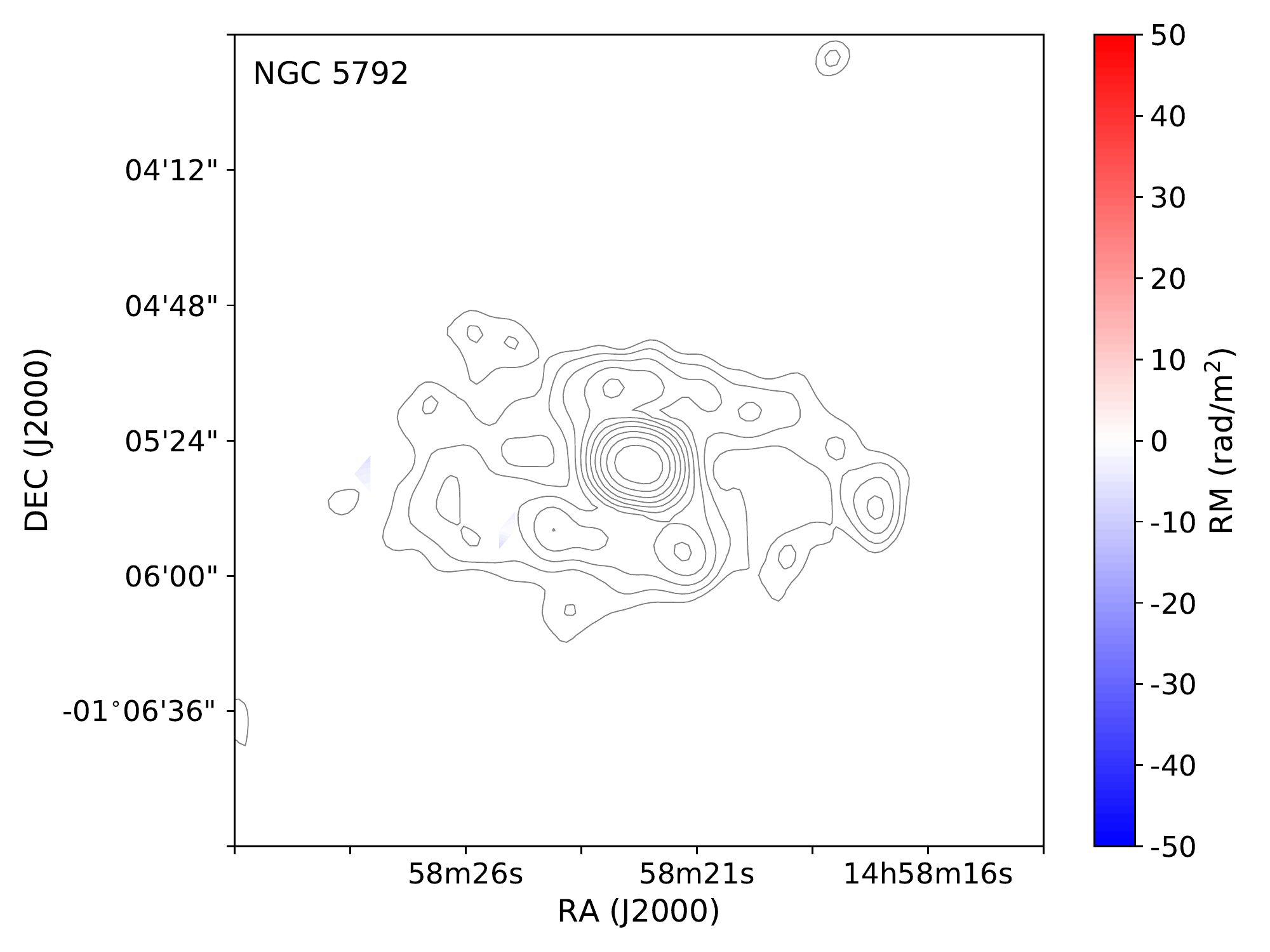}   &
\includegraphics[trim={0 0 0 0}, clip, width=8cm]{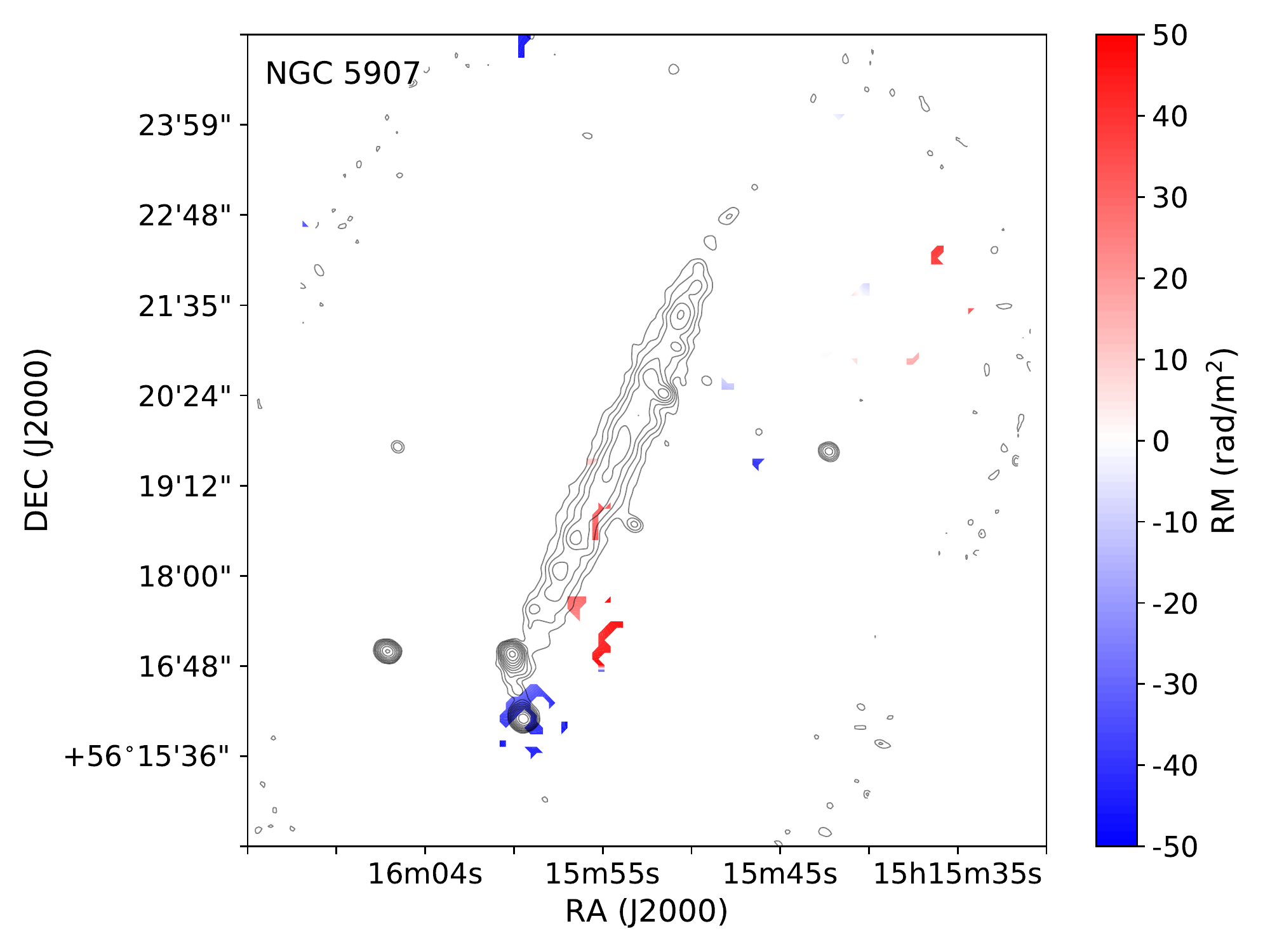}   \\
\end{tabular}
\caption{RM maps (continued)}
\label{fig:RMmaps3}
\end{figure*}

% -----------------------------------------------------------------------

\end{appendix}
% WARNING
%-------------------------------------------------------------------
% Please note that we have included the references to the file aa.dem in
% order to compile it, but we ask you to:
%
% - use BibTeX with the regular commands:
%   \bibliographystyle{aa} % style aa.bst
%   \bibliography{Yourfile} % your references Yourfile.bib
%
% - join the .bib files when you upload your source files
%-------------------------------------------------------------------

\end{document}